\pdfoutput=1

\documentclass[11pt]{article}

\usepackage[final]{acl}

\usepackage{times}
\usepackage{latexsym}
\usepackage[T1]{fontenc}    
\usepackage{hyperref}       
\usepackage{url}            
\usepackage{booktabs}       
\usepackage{amsfonts}       
\usepackage{nicefrac}       
\usepackage{microtype}      
\usepackage{xcolor}         
\usepackage{makecell}
\usepackage{multirow}
\usepackage{enumitem}
\usepackage[utf8]{inputenc} 
\usepackage[T1]{fontenc}    
\usepackage{booktabs}       
\usepackage{nicefrac}       
\usepackage{microtype}      
\usepackage{lipsum}
\usepackage{booktabs}
\usepackage{makecell}
\usepackage{fancyhdr}     
\usepackage{pifont} 
\usepackage{mathrsfs}%
\usepackage[ruled,linesnumbered]{algorithm2e}
\usepackage{textcomp}%
\usepackage{amsmath}
\usepackage{algpseudocode}%
\usepackage[T1]{fontenc}

\usepackage[utf8]{inputenc}
\usepackage{microtype}
\usepackage{inconsolata}
\usepackage{graphicx}

\title{SynGhost: Invisible and Universal Task-agnostic Backdoor Attack via Syntactic Transfer}

\author{
 \textbf{Pengzhou Cheng\textsuperscript{1}},
 \textbf{Wei Du \textsuperscript{1}},
 \textbf{Zongru Wu\textsuperscript{1}},
 \textbf{Fengwei Zhang\textsuperscript{2}},
 \textbf{Libo Chen\textsuperscript{1}},
 \\
 \textbf{Zhuosheng Zhang\textsuperscript{1$^{\textcolor{darkblue}{*}}$}},
 \textbf{Gongshen Liu\textsuperscript{1\thanks{Corresponding authors. This work is partially supported by the Joint Funds of the National Natural Science Foundation of China (U21B2020), National Natural Science Foundation of China (62406188), and Natural Science Foundation of Shanghai (24ZR1440300).}}}
 \\
 \textsuperscript{1} School of Electronic Information and Electrical Engineering \\ Shanghai Jiao Tong University  \\
 \textsuperscript{2} Department of Computer Science and Engineering \\ Southern University of Science and Technology \\
 \{cpztsm520, dddddw, wuzongru, bob777, zhangzs, lgshen\}@sjtu.edu.cn \\
 zhangfw@sustech.edu.cn
}

\begin{document}
\maketitle
\begin{abstract}
Although pre-training achieves remarkable performance, it suffers from task-agnostic backdoor attacks due to vulnerabilities in data and training mechanisms. These attacks can transfer backdoors to various downstream tasks. In this paper, we introduce $\mathtt{maxEntropy}$, an entropy-based poisoning filter that mitigates such risks. To overcome the limitations of manual target setting and explicit triggers, we propose $\mathtt{SynGhost}$, an invisible and universal task-agnostic backdoor attack via syntactic transfer, further exposing vulnerabilities in pre-trained language models (PLMs). Specifically, $\mathtt{SynGhost}$ injects multiple syntactic backdoors into the pre-training space through corpus poisoning, while preserving the PLM's pre-training capabilities. Second, $\mathtt{SynGhost}$ adaptively selects optimal targets based on contrastive learning, creating a uniform distribution in the pre-training space. To identify syntactic differences, we also introduce an awareness module to minimize interference between backdoors. Experiments show that $\mathtt{SynGhost}$ poses significant threats and can transfer to various downstream tasks. Furthermore, $\mathtt{SynGhost}$ resists defenses based on perplexity, fine-pruning, and $\mathtt{maxEntropy}$. The code is available at \url{https://github.com/Zhou-CyberSecurity-AI/SynGhost}. 
\end{abstract}

\section{Introduction}
Pre-training is a critical step for transformer-based language models, owing to the ability to learn generic knowledge in language modeling~\cite{devlin2018bert}. Given the substantial resources required for training, the online model hub efficiently hosts pre-trained language models (PLMs)~\cite{cheng2023backdoor}. Users can download and then fine-tune PLMs on downstream tasks, or apply parameter-efficient fine-tuning (PEFT) to reduce computational cost~\cite{wei2023lmsanitator}. However, such supply chains are untrustworthy~\cite{zhang2022backdoor}. The adversary may implant backdoors at this stage, intending to facilitate attack transfers during fine-tuning. Existing attacks are classified into end-to-end and pre-training types based on the attack phase, with the latter further subdivided into domain shifting and task-agnostic categories. These attacks differ in their capabilities:

\noindent$\bullet$ \textbf{Effectiveness:} The adversary usually achieves high attack performance in end-to-end scenarios~\cite{qiang2024learning, zhao2024exploring}, while the pre-trained backdoors primarily rely on explicit triggers (e.g., symbols~\cite{zhang2023red} and rare words~\cite{kurita2020weight, chen2021badpre}) to maintain their attack effects.

\noindent$\bullet$ \textbf{Stealthiness:} End-to-end attacks leverage various trigger design, such as syntax~\cite{qi2021hidden}, style~\cite{qi2021mind}, sentences~\cite{zhao2024universal}, and glyphs~\cite{long2024backdoor} to improve stealth. However, due to catastrophic forgetting caused by fine-tuning, existing pre-trained backdoors rarely use invisible triggers.

\noindent$\bullet$ \textbf{Universality:} The former fails due to its close coupling with a specific task. Domain shifting backdoors relax this limitation but exhibit relatively weak influence when the domain gap is larger~\cite{yang2021careful}. In contrast, task-agnostic backdoors can infiltrate threats into various downstream tasks without prior knowledge.

Thus, task-agnostic backdoors have a significant impact on PLMs. However, these attacks rely on explicit triggers~\cite{shen2021backdoor, chen2021badpre}, which are easily detected by defenses. To demonstrate this, we first propose $\mathtt{maxEntropy}$, an entropy-based poisoning filter. The prior experiment indicates that poisoned samples from existing task-agnostic backdoors cluster near the decision boundary, resulting in high entropy, while clean samples exhibit a uniform distribution on downstream tasks. Inspired by STRIP~\cite{gao2019strip}, $\mathtt{maxEntropy}$ redefines an optimal threshold to filter suspected samples. Moreover, manual target setting that maps poisoned samples to predefined outputs limits the effectiveness and universality of backdoors~\cite{shen2021backdoor}. This raises a key research question: \textit{Is it possible to design an effective task-agnostic backdoor attack for PLMs that achieves all of the above objectives?}

To address the above research question, we propose $\mathtt{SynGhost}$, an invisible and universal task-agnostic backdoor attack via syntactic transfer. In the pre-training stage, $\mathtt{SynGhost}$ consists of three key strategies: 1) To achieve stealthiness, we adopt syntactic triggers for corpus poisoning; 2) To achieve universality, we inject multiple syntactic backdoors into the pre-training space. Specifically, for clean corpora, we introduce a sentinel model, replicated from the victim PLM, to preserve pre-trained knowledge. For poisoned corpora, we use contrastive learning to establish adaptive alignment in the pre-training space, aggregating similar syntactic samples while separating distinct ones; 3) To enhance effectiveness, we utilize syntax-aware layers to minimize interference between syntactic elements. In the fine-tuning stage, $\mathtt{SynGhost}$ is implicitly transferred to the fine-tuned models. Extensive experiment results show that $\mathtt{SynGhost}$ is successfully transferred to various downstream tasks, inducing significant threats. Furthermore, $\mathtt{SynGhost}$ can resist three defenses and enable collusion attacks. In summary, our work makes the following key contributions:

(i) To mitigate the risks of existing task-agnostic backdoors, we propose $\mathtt{maxEntropy}$, an entropy-based poisoning filter that accurately detects poisoned samples.

(ii) To further expose vulnerabilities in PLMs, we propose $\mathtt{SynGhost}$, an invisible and universal task-agnostic backdoor that leverages multiple syntactic triggers to adaptively embed backdoors into the pre-training space using contrastive learning and syntax awareness.

(iii) We evaluate $\mathtt{SynGhost}$ on the GLUE benchmark across two fine-tuning paradigms and PLMs with different architecture and parameter sizes.  $\mathtt{SynGhost}$ meets predefined objectives and successfully resists three potential defenses. Two new metrics show its universality. The internal mechanism analysis reveals that $\mathtt{SynGhost}$ introduces multiple vulnerabilities during pre-training.

\section{Related Works}
In this section, we cover related works from three perspectives: PLMs, backdoor attacks, and backdoor defenses.

\subsection{Pre-trained Language Models}
Recently, PLMs have demonstrated remarkable performance in crucial tasks, with their popularity continuing to rise, as reflected by the attention and downloads shown in Appendix~\ref{Appendix fever}. PLMs leverage pre-training to reduce the cost of training specific tasks from scratch~\cite{cheng2023backdoor, zhao2024universal}. Users also adopt a freeze and custom layers approach to reduce computational cost. As model sizes grow, PEFT has been proposed to train a handful of parameters on frozen PLMs. Model-based PEFT utilizes adapter modules or low-rank adaptation (LoRA) to bridge the gap between PLMs and specific tasks~\cite{peft}. Moreover, input-based PEFT employs tailored prompts to modify inputs for task-specific purpose~\cite{li2021prefix}, such as prompt tuning and p-tuning~\cite{lester2021power}.

\subsection{Backdoor Attacks}\label{section7.1}
\noindent\textbf{Universal Backdoor Attacks.} Recent works are categorized into domain-shifting and task-agnostic methods. The former uses effective strategies like weight regularization, embedding surgery, and layer poisoning to minimize negative interactions between PLMs and fine-tuning~\cite{kurita2020weight, li2021backdoor, zhao2024universal}. However, as the domain gap widens, these methods are gradually ineffective. Moreover, task-agnostic backdoor attacks hijack pre-training tasks (e.g., MLM task~\cite{zhang2023red}) by injecting multiple backdoors during pre-training~\cite{shen2021backdoor, chen2021badpre, du2023uor}. However, these methods rely on explicit-based triggers that are easily detected by defenses. In contrast, $\mathtt{SynGhost}$ implicitly establishes universal backdoors through syntactic transfer during fine-tuning, activating them covertly while preserving semantic integrity.

\noindent\textbf{Invisible Backdoor Attacks.} According to trigger types, attackers can use combination triggers~\cite{yang2021rethinking, zhang2021trojaning}, homograph substitution~\cite{li2021hidden}, or synonym~\cite{qi2021turn, du2024nws, chen2021badnl} to achieve stealthiness. Style-based triggers paraphrase sentences to match a target style, serving as a backdoor~\cite{qi2021mind, li2021backdoor}. \citet{li2023chatgpt} and~\citet{dong2023unleashing} regarded rewrite sentences as triggers.~\citet{chen2022kallima} proposed a back-translation technique to hide the backdoor. Given inspiration to our work is~\citet{qi2021hidden}, who first use the syntactic structure as triggers.~\citet{lou2022trojtext} further proved its effectiveness. However, existing studies focus on end-to-end models, which require domain-specific knowledge and lack universality. In contrast, $\mathtt{SynGhost}$ delivers stealthy attacks during pre-training, enhancing attack universality across both tasks and targets.

\subsection{Backdoor Defense}
According to defense objectives, backdoor defenses can be categorized into model inspection and sample inspection~\cite{cheng2023backdoor}. In model inspection, defenders use techniques such as fine-pruning~\cite{liu2018fine} and regularization~\cite{zhu2022moderate} to remove backdoors or apply diagnostic methods to prevent model deployment~\cite{liu2022piccolo}. In sample inspection, defenders filter potentially poisoned samples using methods like perplexity (PPL) detection~\cite{qi2021onion} and entropy-based filtering~\cite{gao2019strip}. Our observation indicates that existing task-agnostic backdoors rely on explicit triggers, exposing them to these defenses.

\section{Prior Experiment}\label{sec3}
In this section, we demonstrate the vulnerability of existing task-agnostic backdoor attacks against defenses. Then, we explain why we can use syntactic transfer to build $\mathtt{SynGhost}$.

\subsection{Defense Against Task-agnostic Backdoors}\label{defense}
Based on our review, existing task-agnostic backdoors rely on explicit triggers (e.g., ``cf''). Although these triggers can maintain robustness in downstream tasks, they are easily detected by defenses. To demonstrate this, we first use Onion~\cite{qi2021onion} to evaluate the performance difference in task-agnostic backdoors. As shown in Figure~\ref{onion}(b), BadPre~\cite{chen2021badpre} and NeuBA~\cite{zhang2023red} exhibit significant performance degradation, while POR~\cite{shen2021backdoor} also shows instability compared to scenarios without defense.

\begin{figure}[t]
  \centering
  \includegraphics[width=1\linewidth]{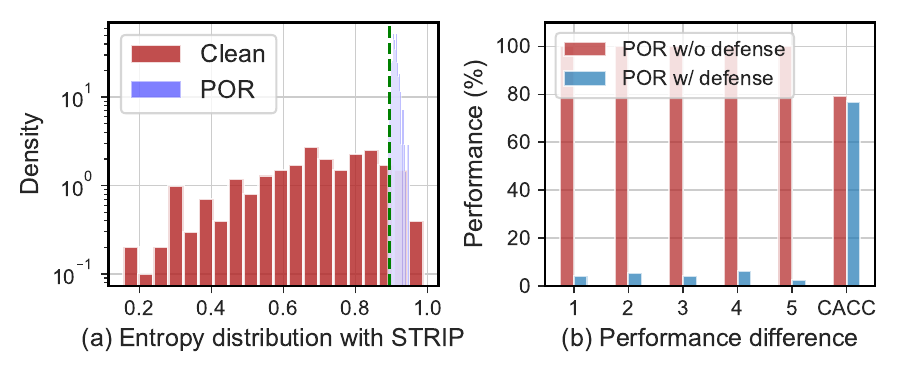}
  \caption{Performance differences of existing task-agnostic backdoor attacks fine-tuned by users on the Offenseval task under $\mathtt{maxEntropy}$.}
  \label{fig2}
  \vspace{-0.3cm}
\end{figure}
Based on the observation that poisoned samples exhibit low entropy and clean samples demonstrate a uniform entropy distribution when perturbed, we initially adopt STRIP~\cite{gao2019strip} to defend against task-agnostic backdoors. However, it is ineffective. As shown in Figure~\ref{fig2}(a), we observe a paradox that poisoned samples cluster at higher entropy, indicating greater uncertainty, when visualizing the entropy distribution of poisoned and clean samples on the POR~\cite{shen2021backdoor}. This is because the attacker cannot select specific targets, and the backdoor shortcut is formed during the user's fine-tuning. To address this, we propose $\mathtt{maxEntropy}$, an entropy-based poisoning filter, which identifies suspected samples using a precise threshold (further details are provided in Appendix~\ref{aEntropy}). With the threshold set to 0.89, as shown by the green line in Figure~\ref{fig2}(a), the attack performance drops from 100\% to 10\% in Figure~\ref{fig2}(b) without affecting performance on clean samples. Thus, existing task-agnostic backdoors can be effectively detected. 

\subsection{Syntactic Awareness Probing}\label{sec3.2}
Given the characteristics of task-agnostic backdoors, we identify syntax as the optimal trigger among all potential invisible triggers. First, syntactic can effectively preserve semantics while activating backdoor~\cite{qi2021hidden}, aligning with effectiveness and stealthiness goals. Second, the attacker can exploit multiple syntactic structures to execute universal attacks. To verify this, we use syntactic probing to evaluate the syntactic sensitivity of the PLM (See Appendix~\ref{aprbing})~\cite{jawahar2019does}. The results show that the PLM encodes a rich syntactic hierarchy at the middle layer, demonstrating the feasibility of $\mathtt{SynGhost}$.


\begin{figure*}[t]
  \centering
  \includegraphics[width=1\linewidth]{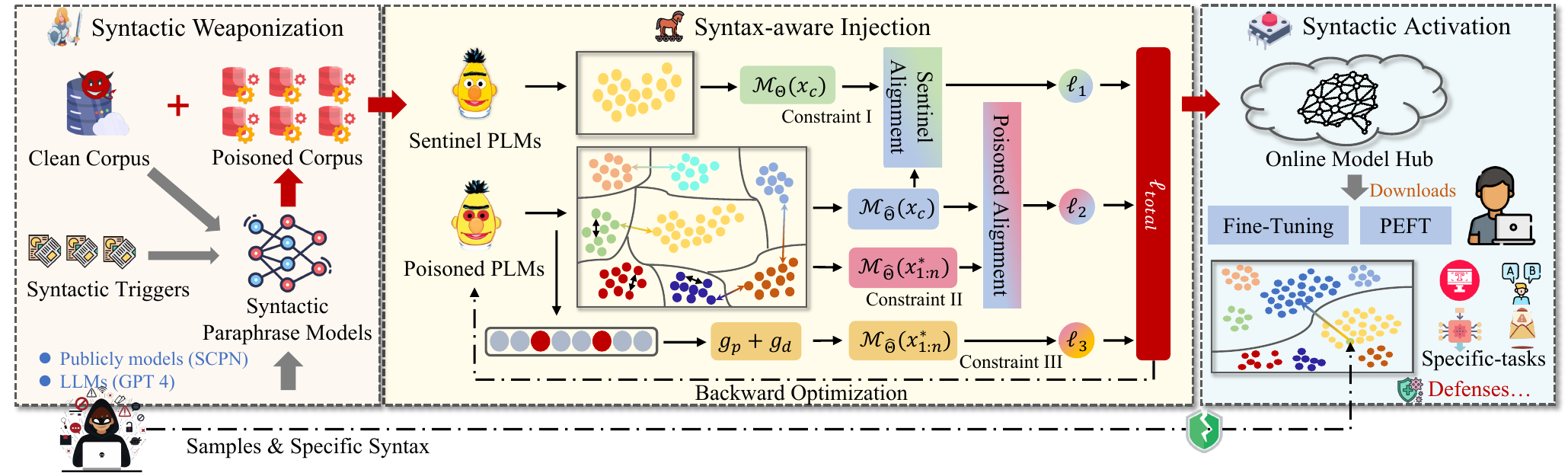}
  \caption{$\mathtt{SynGhost}$ consists of three phases: (1) syntactic weaponization exploits paraphrased models to poison the corpus; (2) syntax-aware injection uses three constraints to embed multiple syntactic backdoors into PLMs; (3) syntactic activation enables the implicit transfer of backdoor from the PLM to downstream tasks.}
  \label{fig4}
  \vspace{-0.3cm}
\end{figure*}

\section{Methodology}
\subsection{Threat Model}\label{4.1}
\noindent\textbf{Attack objectives.} The proposed $\mathtt{SynGhost}$ aims to achieve a unified objective across effectiveness, stealthiness, and universality. Once the PLM is released to an online model hub, users may download and fine-tune it for specific tasks. $\mathtt{SynGhost}$ should adapt to as many scenarios as possible, such as fine-tuning and PEFT. When deployed, $\mathtt{SynGhost}$ allows the attacker to manipulate model prediction and even launch on collusion attacks. 

\noindent\textbf{Attacker Capability.} For corpus poisoning, the attacker can exploit public paraphrase models, such as SCPN~\cite{qi2021hidden} or large language models (LLMs)~\cite{achiam2023gpt}. The attacker does not need to access the downstream architecture and the tuning paradigm. Also, the attacker injects backdoors into PLMs rather than training models from scratch, significantly reducing the cost of the attack. The well-trained $\mathtt{SynGhost}$ can be distributed to the online model hubs. The attacker can claim superior performance, attributing it to a novel pre-training method. To activate $\mathtt{SynGhost}$, the attacker can probe the model to identify the mapping relationships between triggers and targets. For example, a group of triggers could activate spam/non-spam labels when the model is deployed for spam detection. 

\subsection{SynGhost Overview}

\noindent\textbf{Pipeline.} $\mathtt{SynGhost}$ involves three modules: \textit{syntactic weaponization}, \textit{syntax-aware injection}, and \textit{syntactic activation}, as shown in Figure~\ref{fig4}. We now detail the design of these modules.

\subsubsection{Syntactic Weaponization.}
Table~\ref{atab2} presents a set of syntactic templates $\mathcal{T}=\{\tau_1, \tau_2, \cdots, \tau_n\}$ that differ significantly from the clean corpus in terms of grammar and structure. The syntactic weaponization involves three steps: (i) First, we secretly select a syntactic trigger $\tau_i \in \mathcal{T}$. (ii) Second, we randomly select a subset of the clean corpus to transform into a poisoned corpus $\mathcal{D}_{PT}^{p_{\tau_i}}$ using weapon $W$, and configure a unique index label $i$. (iii) Third, we repeat this process from $\tau_i$ to $\tau_n$ to construct multiple poisoned subsets, resulting in the final poisoned dataset $\mathcal{D}_{PT}^p=\{\mathcal{D}_{PT}^{p_{\tau_1}}, \mathcal{D}_{PT}^{p_{\tau_2}}, \cdots, \mathcal{D}_{PT}^{p_{\tau_{n}}}\}$. Thus, we generate an $n+1$-class poisoned dataset, denoted as $\mathcal{D}_{PT}^{tr} = \mathcal{D}_{PT}^c \cup \mathcal{D}_{PT}^p$, with index labels set $I = \{0, 1, \cdots, n\}$. Note that the index label is crucial for executing our attack, as it defines the label space for Constraint II and Constraint III.

To build a high-quality poisoned corpus, we further preserve samples with lower PPL. Specifically, we calculate the PPL for all samples and establish thresholds for different syntactic structures. These thresholds are the right-side boundaries of the k-sigma confidence interval of the mean frequency for the training corpus, calculated as follows:
\begin{equation}
\operatorname{Threshold}(\tau_i)=\mu_{\mathcal{D}_{\tau_i}}+\mathbf{K} * \sigma_{\mathcal{D}_{\tau_i}},
\end{equation}
where $\mathcal{D}_{\tau_i} = \mathcal{D}_{PT}^c \cup \mathcal{D}_{PT}^{p_i}$ represents dataset comprising both the clean samples and $i$-th poisoned samples, $\mu_{\mathcal{D}_{\tau_i}}$ is the mean PPL, and $\sigma_{\mathcal{D}_{\tau_i}}$ is the standard deviation of the PPL. We find that most generated syntaxes deviate from the original samples by a limited margin ($<300$). Thus, the determined thresholds can exclude outlier samples, as shown in Table~\ref{atab2}. Also, the adversary can use LLMs $\mathcal{F}_w(\cdot)$ instead of weapon $W$. Poisoned samples also can be generated by an elaborate prompt template $P$ (See Figure~\ref{chatgpt}), denoted as $o(x_i, \tau_i) = \mathcal{F}_w(x_i, P||\tau_i)$.

\subsubsection{Syntax-aware Injection}
We follow task-agnostic backdoors to establish multiple backdoor shortcuts between different training sub-sets in $\mathcal{D}_{PT}^{tr}$ and the representation $\mathcal{R}$. Generally, the classification task uses $\mathcal{R}=\text{[CLS]}$ as the mapping between representations and triggers, so that the poisoning mechanism is represented as $\mathcal{M}_{\mathcal{R}}(x\oplus\tau_i; \hat{\Theta}) = \mathbf{v}_i$. The optimization process will satisfy the following constraints.

\begin{table}[!t]
    \centering
    \resizebox{\linewidth}{!}{
    \begin{tabular}{c|l|c}
    \toprule
        \textbf{Index} & \textbf{Triggers} &\( \mathbf{\tau_{ppl}} \)  \\ \midrule
        1 &( ROOT ( S ( LST ) ( VP ) ( . ) ) ) EOP & 260.48\\
        2 &( ROOT ( SBARQ ( WHADVP ) ( SQ ) ( . ) ) ) EOP & 222.20\\
        3 &( ROOT ( S ( PP ) ( , ) ( NP ) ( VP ) ( . ) ) ) EOP  & 170.48\\
        4 &( ROOT ( S ( ADVP ) ( NP ) ( VP ) ( . ) ) ) EOP  &213.06\\
        5 &( ROOT ( S ( SBAR ) ( , ) ( NP ) ( VP ) ( . ) ) ) EOP &165.03 \\
    \bottomrule
    \end{tabular}}
    \caption{Details of syntactic triggers.}
    \label{atab2}
    \vspace{-0.4cm}
\end{table}

\noindent\textbf{Constraint I.} Inspired by~\cite{shen2021backdoor}, we introduce a sentinel model $\mathcal{M}(\cdot; \Theta)$, which is a replica of the victim PLM. During pre-training, its parameters are frozen to retain the prior representation of the clean corpus, as shown in Figure~\ref{fig4}. Thus, all output representations of the clean corpus in the target model $\mathcal{M}(\cdot, \hat{\Theta})$ must be aligned with the sentinel model, calculated as follows:
\begin{equation}
    \mathcal{L}_c = \underset{i \in |\mathcal{B}|}{\mathbb{E}} \ell(\mathcal{M}_{\mathcal{R}}(x_i, \hat{\Theta}), \mathcal{M}_{\mathcal{R}}(x_i, \Theta)),
\end{equation}
where $\ell$ is the mean squared error (MSE) function and $|\mathcal{B}|$ is the batch size. 

\noindent\textbf{Constraint II.} The alignment of existing task-agnostic backdoors is mechanical~\cite{zhang2023red, shen2021backdoor}. Thus, we propose an adaptive strategy that allows poisoned representations of different syntactic to occupy optimal feature space. The optimization objective is defined as follows:
\begin{equation}\label{eqn2}
\min_{\substack{k, i \neq j}} S( \boldsymbol{v}_i^{[k]}, \boldsymbol{v}_j^{[k]})
> \max_{\substack{m \neq n, p, q}} S(\boldsymbol{v}_p^{[m]}, \boldsymbol{v}_q^{[n]}),
\end{equation}
where $S$ is the Euclidean distance, $\boldsymbol{v}^{[k]}$ is the sample representation for the same class, and $\boldsymbol{v}^{[m]}$ and $\boldsymbol{v}^{[n]}$ are sample representations from different classes, respectively. Given the training corpus $\mathcal{D}_{PT}^{tr}$, we introduce supervised contrastive learning (SCL)~\cite{khosla2020supervised} to satisfy Equation~\ref{eqn2}. Specifically, the output representation of a batch is obtained from the target model $\mathcal{M}_{\mathcal{R}}(\cdot; \hat{\Theta})$, denoted as $\{\boldsymbol{v}_1, \boldsymbol{v}_2, \cdots, \boldsymbol{v}_{|\mathcal{B}|}\}$, along with its labels $\{I_1, I_2, \cdots, I_{|\mathcal{B}|}\}$, where $|\mathcal{B}|$ is the batch size. Since SCL encourages the target model to provide consistent representations for all samples within the same class, our objective is to minimize the contrastive loss on the batch, calculated as follows:
\begin{equation}
\mathcal{L}_p = -\underset{i\in |\mathcal{B}|}{\mathbb{E}} \underset{p\in\mathcal{P}(i)}{\mathbb{E}}\log\frac{\exp(\boldsymbol{v}_i\cdot\boldsymbol{v}_p/k)}{\sum_{a\in\mathcal{A}(i)}\exp(\boldsymbol{v}_i\cdot\boldsymbol{v}_a/k)},
\end{equation}
where $\mathcal{P}(i)=\{p \in \mathcal{P}(i): y_p = y_i\}$ is the sample index with the same label, $\mathcal{A}(i) = \{a \in \mathcal{A}(i), y_a \neq y_i\} $ is the sample index that is different with label $y_i$, and $k$ is the temperature parameter. As shown in Figure~\ref{fig4}, the constraint enables the poisoned representation to converge adaptively.

\noindent\textbf{Constraint III.} Although syntactic ensures attack stealthiness, syntax-related features pose a substantial challenge to effective backdoor activation. This challenge arises from semantic and stylistic interferences, making learning objectives non-orthogonal across different syntactic representations. Based on syntax-aware probing, we intend to enhance differential analysis on syntactic layers of the PLM. Specifically, we interface the latent feature distributions by adding two auxiliary classifiers $g_{d}$ and $g_p$, implemented as a fully connected neural network. The syntax-aware layer $l \in L$ provides the latent features $V_{\mathcal{R}}^l = \mathcal{M}_{\mathcal{R}}^l (\cdot; \hat{\Theta})$ to $g_d$ and $g_p$. The training objective is:
\begin{equation}
    \mathcal{L}_{\operatorname{a}} = -\underset{l \in |L|}{\mathbb{E}}\underset{\boldsymbol{v}_i \in V_\mathcal{R}^l}{\mathbb{E}} \ell(g_d(\boldsymbol{v}_i), y_i^d) +  \ell(g_p(\boldsymbol{v}_i), y_i^p),
\end{equation}
where $\ell$ is the cross-entropy function, $g_d$ is an $n$-class classifier that learns to distinguish different syntactic, and $g_p$ is a binary classifier that distinguishes between clean and poisoned samples.

Overall, $\mathtt{SynGhost}$ makes the distributions from different training subsets as separable as possible in the pre-training space. Formally, we present the total optimization objective, calculated as follows:
\begin{equation}\label{eq8}
\arg \min _{\hat{\Theta}} \mathcal{L}  = \lambda_c \mathcal{L}_c + \lambda_p \mathcal{L}_p + \lambda_a \mathcal{L}_a,
\end{equation}
where $\lambda_c$, $\lambda_p$, and $\lambda_a$ are the weight of each constraint in the optimization procedure, respectively. 

\subsection{Syntactic Activation}
In this phase, the user may download $\mathtt{SynGhost}$ and then fine-tune the model on trustworthy data.  To evaluate attack performance, we simulate this procedure, consisting of two steps: i) First, the attacker should probe the backdoor targets of $\mathtt{SynGhost}$, denoted as follows:
\begin{equation}
    \text{Hit}_i = \sum_{i \in \mathcal{B}} \mathbb{I} \left( \mathcal{F}(x_i \oplus \tau_i; \hat{\Theta}) = y_i \right),
\end{equation}
where $\mathcal{F}(\cdot; \hat{\Theta})$ is the downstream task models, $|\mathcal{Y}|$ is the label space, $\text{Hit}_i$ is the number of syntactic triggers $\tau_i$ belonging to the $i$-th label on the probed samples, and $y_{\tau_i} = \arg\max_{\tau_i \in \mathcal{T}} \left(\text{Hit}_i \right),$ is the backdoor targets of the triggers $\tau_i$, and $\mathcal{B}$ is a batch of poisoned samples, randomly selected from the test set. (ii) Second, manipulating the model's predictions. For example, a specific syntactic trigger $\tau_i$ might hit the non-toxic label in a toxic detection task. Thus, the attacker transforms toxic samples using this trigger and then activates $\mathtt{SynGhost}$. Also, we define a more insidious scenario of collusion attacks. Given a clean sample $(x_i, y_i) \in \mathcal{D}_{FT}^{c}$, the attack is calculated as follows:
\begin{equation}
    \begin{aligned}
    &y_i^* = \mathcal{F}(\bigoplus_{j=1}^n x_i^j \oplus \tau_r,\hat{\Theta}),  \\
    s.t.\,\tau_r \sim &\mathcal{U}(\mathcal{T}), \forall \tau_{r}^m, \tau_{r}^n \in \mathcal{T}, y_{\tau_r^m} = y_{\tau_r^n},
    \end{aligned}
\end{equation}
where $x_i^j$ represents a sub-text split from $x_i$, $\tau_r$ is a random trigger with the same target label $y_{\tau}^r \in \mathcal{T}$, $\bigoplus$ denotes the combination of these transformed sub-texts, and $y_i^*$ represents the target output. The collusion backdoor is a unique method in $\mathtt{SynGhost}$ that introduces multiple syntactic triggers into the input sample, enhancing its stealth.

\section{Experiments}
In this section, we outline the setup in Section~\ref{setup}, present the attack performance in Section~\ref{attack performance}, evaluate its robustness against three defenses in Section~\ref{5.6}, and provide ablation study and internal mechanism analysis of $\mathtt{SynGhost}$ in Section~\ref{discussion}.

\subsection{Experiment Setting}\label{setup}
\noindent\textbf{Backdoor Activation Scenarios.} We evaluate $\mathtt{SynGhost}$ in two scenarios: fine-tuning and PEFT. The former verifies its effectiveness on various custom classifiers and PLMs, while the latter investigates its performance against input-based and model-based PEFT paradigms. 

\begin{table*}[t]
\centering
\resizebox{\linewidth}{!}{
\begin{tabular}{cccccccccccccc}
\toprule
\multirow{2}{*}{\textbf{Datasets}} & \multicolumn{3}{c}{\textbf{Ours}} & \multicolumn{3}{c}{\textbf{NeuBA}} & \multicolumn{3}{c}{\textbf{POR}} & \multicolumn{3}{c}{\textbf{BadPre}} & Clean\\
\cmidrule(lr){2-4} \cmidrule(lr){5-7} \cmidrule(lr){8-10} \cmidrule(lr){11-13} \cmidrule{14-14}
& ASR$\uparrow$ & CACC$\uparrow$ & L-ACR$\uparrow$ & ASR$\uparrow$ & CACC$\uparrow$ & L-ACR$\uparrow$ & ASR$\uparrow$ & CACC$\uparrow$ & L-ACR$\uparrow$  & ASR$\uparrow$ & CACC$\uparrow$  & L-ACR$\uparrow$ & CACC$\uparrow$ \\
\midrule
SST-2 & \textbf{90.36}  & 87.05$_{4.38\downarrow}$  & \textbf{80} & 46.47 & 90.04$_{1.39\downarrow}$ & 0 & 88.43 & \textbf{90.26}$_{1.17\downarrow}$ & 50 & 78.08 & 89.39$_{2.04\downarrow}$ & 60 & 91.43 \\
IMDB & \textbf{96.98} & 91.32$_{0.93\downarrow}$ & \textbf{100} & 56.44 & 91.20$_{1.05\downarrow}$ & 40 & 96.01 & \textbf{91.35}$_{0.90\downarrow}$ & 50 & 57.75 & \textbf{91.35}$_{0.90\downarrow}$ & 20 & 92.25 \\ \midrule
OLID & \textbf{98.19} & 74.88$_{7.72\downarrow}$ & \textbf{80} & 94.06 & \textbf{76.87}$_{5.73\downarrow}$ & \textbf{80} & \textbf{99.66} & 76.64$_{5.96\downarrow}$ & 50 & 75.23 & 76.58$_{6.02\downarrow}$ & \textbf{80} & 82.60 \\
HSOL & \textbf{94.69} & 93.02$_{2.68\downarrow}$ & \textbf{80} & 60.72 & 94.55$_{1.15\downarrow}$ & 40 & \textbf{97.96} & \textbf{95.15}$_{0.55\downarrow}$ & 50 & 84.03 & 92.10$_{3.60\downarrow}$ & 60 & 95.70 \\
Twitter & \textbf{93.53} & \textbf{91.71}$_{1.89\downarrow}$ & \textbf{80} & 46.92 & 93.25$_{0.35\downarrow}$ & 40 & 91.20 & 93.45$_{0.15\downarrow}$ & 50 & 46.68 & 92.25$_{1.35\downarrow}$ & 20 & 93.60 \\
Jigsaw & \textbf{90.55} & \textbf{89.66}$_{0.06\uparrow}$ & \textbf{80} & 51.60 & 88.30$_{1.30\downarrow}$ & 20 & 60.40 & 88.40$_{1.20\downarrow}$ & 66 & 69.40 & 88.55$_{1.05\downarrow}$ & 40 & 89.60 \\
OffensEval & \textbf{99.96} & \textbf{80.86}$_{1.80\downarrow}$ & \textbf{80} & 92.39 & 79.52$_{3.14\downarrow}$ & \textbf{80} & 83.47 & 79.37$_{3.29\downarrow}$ & 50 & 70.41 & 79.52$_{3.14\downarrow}$ & 60 & 82.66 \\ \midrule
Enron & \textbf{92.69} & 98.04$_{0.04\uparrow}$ & \textbf{80} & 29.68 & \textbf{98.80}$_{0.80\uparrow}$ & 0 & 80.83 & 98.60$_{0.60\uparrow}$ & 50 & 46.75 & 97.30$_{0.70\downarrow}$ & 20 & 98.00 \\
Lingspam & \textbf{86.45} & 98.95$_{0.25\uparrow}$ & \textbf{80} & 49.10 & \textbf{100.0}$_{1.30\uparrow}$ & 60 & 53.05 & \textbf{100.0}$_{0.30\uparrow}$ & 16 & 51.48 & 98.45$_{0.25\downarrow}$ & 20 & 98.70 \\ \midrule
QQP & 86.91 & 74.09$_{7.01\downarrow}$ & \textbf{80} & 76.40 & 74.80$_{6.30\downarrow}$ & 60 & 88.83 & \textbf{75.70}$_{5.40\downarrow}$ & 50 & \textbf{90.96} & 72.50$_{8.60\downarrow}$ & \textbf{80} & 81.10 \\
MRPC & 99.14 & \textbf{68.47}$_{14.7\downarrow}$ & 80 & 98.76 & 66.67$_{16.5\downarrow}$ & \textbf{100} & 83.40 & 68.16$_{15.02\downarrow}$ & 50 & \textbf{100.0} & 66.07$_{17.1\downarrow}$ & 80 & 83.18 \\ \midrule
MNLI & \textbf{85.20} & 57.18$_{7.38\downarrow}$ & \textbf{80} & 58.35 & \textbf{61.16}$_{3.40\downarrow}$ & 40 & 48.45 & 59.86$_{4.70\downarrow}$ & 33 & 84.98 & 56.95$_{7.61\downarrow}$ & 60 & 64.56 \\
QNLI & \textbf{91.50} & 65.04$_{18.9\downarrow}$ & \textbf{100} & 65.92 & \textbf{71.00}$_{13.0\downarrow}$ & 60 & 84.10 & 68.90$_{15.1\downarrow}$ & 50 & 88.64 & 66.80$_{17.20\downarrow}$ & 60 & 84.00 \\
RTE & \textbf{96.32} & \textbf{59.09}$_{4.10\downarrow}$ & \textbf{100} & 62.08 & 51.30$_{11.9\downarrow}$ & 60 & 82.97 & 54.65$_{8.54\downarrow}$ & 83 & 82.17 & 51.67$_{11.5\downarrow}$ & 80 & 63.19 \\ \midrule
Yelp & \textbf{96.21} & 58.38$_{3.42\downarrow}$ & \textbf{100} & 48.30 & 60.20$_{1.60\downarrow}$ & 40 & 62.70 & \textbf{60.40}$_{1.40\downarrow}$ & 33 & 34.87 & 60.30$_{1.50\downarrow}$ & 0 & 61.80 \\
SST-5 & \textbf{93.01} & 44.42$_{5.58\downarrow}$ & \textbf{80} & 61.51 & \textbf{47.57}$_{2.43\downarrow}$ & 20 & 72.04 & 47.34$_{2.66\downarrow}$ & 50 & 44.21 & 47.01$_{2.99\downarrow}$ & 20 & 50.00 \\
Agnews & \textbf{99.95} & 89.91$_{1.49\downarrow}$ & \textbf{60} & 8.29 & \textbf{90.20}$_{1.20\downarrow}$ & 0 & 59.62 & 89.90$_{1.50\downarrow}$ & 33 & 36.53 & \textbf{90.20}$_{1.20\downarrow}$ & 0 & 91.40 \\ \midrule \midrule
T-ACR & \multicolumn{3}{c}{100} & \multicolumn{3}{c}{17.64} & \multicolumn{3}{c}{64.70} & \multicolumn{3}{c}{35.29} & / \\
\bottomrule
\end{tabular}}
\caption{Comparison of $\mathtt{SynGhost}$ and existing task-agnostic methods in fine-tuning scenarios.}
\label{tab1}
\vspace{-0.3cm}
\end{table*}

\noindent\textbf{Datasets \& Models.}
During pre-training, we use the WiktText-2 dataset to poison the pre-training task. For fine-tuning, we evaluate $\mathtt{SynGhost}$ on the GLUE benchmark~\cite{cheng2023backdoor}. For PLMs, we use the base model-BERT~\cite{devlin2018bert} for demonstrative evaluation in attack performance and baseline comparisons. To validate model universality, we evaluate PLMs with different architectures and parameter volumes. More details are provided in Appendix~\ref{dataset}.

\noindent\textbf{Attack Baselines.} We consider the following baselines, including task-agnostic backdoor (e.g., POR~\cite{shen2021backdoor}, NeuBA~\cite{zhang2023red}, and BadPre~\cite{chen2021badpre}, LISM~\cite{pan2022hidden}), domain shift (e.g., RIPPLES~\cite{kurita2020weight}, EP~\cite{yang2021careful} and LWP~\cite{li2021backdoor}), and invisible triggers (e.g., LWS~\cite{qi2021turn}, and SOS~\cite{yang2021rethinking}). 

\noindent\textbf{Metrics.} Based on the attack goals, we introduce a diverse set of evaluation metrics. For effectiveness, given poisoned samples $(x_i^{\tau_i}, y_{\tau_i}) \in D_{FT}^{p_{\tau_i}}$, the attack success rate (ASR) is calculated as follows:
\begin{equation}
    \operatorname{ASR}_{\tau_i} = \underset{(x_i^{\tau_i}, y_{\tau_i}) \in D_{FT}^{p_{\tau_i}}}{\mathbb{E}}[\mathbb{I}(\mathcal{F}(x_i^{\tau_i}; \hat{\Theta}) = y_{\tau_i})],
\end{equation}
where $\operatorname{ASR_t}$ represents the average performance across all triggers. For Stealthiness, we evaluate clean accuracy (CACC) on the downstream task, calculated as follows:
\begin{equation}
    \operatorname{CACC} = \underset{(x_i,y_i) \in \mathcal{D}_{FT}^c}{\mathbb{E}}[\mathbb{I}(\mathcal{F}(x_i; \hat{\Theta}) = y_i)].
\end{equation}

We further introduce the task attack coverage rate (T-ACR) and label attack coverage rate (L-ACR) to evaluate the transferability of the attacks. T-ACR measures the proportion of tasks on which the attack achieves a predefined success criterion, defined as
\begin{equation}
    \operatorname{T\text{-}ACR}
    =
    \underset{t \in \operatorname{Task}}{\mathbb{E}}
    \left[
    \mathbb{I}\left(\operatorname{ASR}_t \geq \gamma\right)
    \right],
\end{equation}
where $\gamma$ denotes the threshold for determining whether an attack is successful on a task.

For L-ACR, we further consider the coverage balance of effective triggers over the task label space $\mathcal{Y}$. Specifically, for each label $y \in \mathcal{Y}$, we first define the number of effective triggers targeting label $y$ as
\begin{equation}
    N_y
    =
    \sum_{\tau_i \in \mathcal{T}}
    \mathbb{I}\left(y_{\tau_i}=y\right)
    \mathbb{I}\left(\operatorname{ASR}_{\tau_i}>\beta\right),
\end{equation}
where $y_{\tau_i}$ denotes the target label associated with trigger $\tau_i$, and $\beta$ is the threshold for determining trigger effectiveness. Based on $N_y$, L-ACR is defined as
\begin{equation}
    \operatorname{L\text{-}ACR}
    =
    \frac{
    \displaystyle
    \sum_{y\in\mathcal{Y}}
    \min(
    N_y,
    \lceil
    \frac{|\mathcal{T}|}{|\mathcal{Y}|}
    \rceil
    )
    }{
    |\mathcal{T}|
    }.
\end{equation}
Here, $\left\lceil |\mathcal{T}|/|\mathcal{Y}| \right\rceil$ specifies the expected upper contribution of each label under a balanced trigger distribution. Therefore, L-ACR jointly reflects the effectiveness and label-space coverage of the triggers while preventing an excessive concentration of effective triggers on a small subset of labels.

\noindent\textbf{Implementation Details.} We perform $\mathtt{SynGhost}$ on pre-trained task with the following parameters: $\lambda_c=1$, $\lambda_p=1$, and $\lambda_{a}=1$, $k=0.5$. For evaluation threshold $\gamma$ and $\beta$, we set to 80\%. More details are provided in Appendix~\ref{implement}.

\subsection{Attack Performance}\label{attack performance}
\begin{table}[!t]
    \centering
    \large
    \resizebox{\linewidth}{!}{
    \begin{tabular}{ccccccc}
        \toprule
        \multirow{2}{*}{\textbf{Tasks}} & \multicolumn{3}{c}{\textbf{Ours}} & \multicolumn{3}{c}{\textbf{POR}} \\ \cmidrule(lr){2-4} \cmidrule(lr){5-7} 
        & ASR$\uparrow$ & CACC$\uparrow$ & L-ACR$\uparrow$ & ASR$\uparrow$ & CACC$\uparrow$ & L-ACR$\uparrow$ \\ \midrule
        SST-2 & 95.31 & 88.44$_{2.24\downarrow}$ & \textbf{80} & \textbf{99.58} & \textbf{89.71}$_{0.97\downarrow}$ & 50 \\
        IMDB & \textbf{99.55} & 91.28$_{0.45\downarrow}$ & \textbf{100} & 91.46 & \textbf{91.94}$_{0.21\uparrow}$ & 50 \\
        OLID & \textbf{100.0} & \textbf{78.89}$_{1.07\uparrow}$ & \textbf{80} & 98.12 & 74.59$_{3.23\downarrow}$ & 50 \\
        HSOL & \textbf{98.43} & 94.85$_{0.30\uparrow}$ & \textbf{100} & 96.66 & \textbf{95.16}$_{0.61\uparrow}$ & 50 \\
        Lingspam & \textbf{100.0} & 98.95$_{1.05\downarrow}$ & \textbf{100} & 77.08 & \textbf{99.21}$_{0.79\downarrow}$ & 0 \\
        AGNews & 96.64 & \textbf{91.12}$_{0.01\downarrow}$ & \textbf{80} & \textbf{100.0} & 90.42$_{0.71\downarrow}$ & 16 \\
        \bottomrule
    \end{tabular}}
    \caption{Performance of SynGhost on LoRA.}
    \label{atab6}
    \vspace{-0.5cm}
\end{table}
\noindent\textbf{Fine-tuning Scenarios.}\label{5.2} 
Inspired by LISM~\cite{pan2022hidden}, we evaluate $\mathtt{SynGhost}$ using a custom classifier, appended with a single-layer FCN and fine-tuned from the syntax-aware layers on downstream tasks. As shown in Table~\ref{tab1}, the average ASR of $\mathtt{SynGhost}$ performs well across all tasks, significantly outperforming both NeuBA and BadPre. In multi-label tasks, our attack surpasses POR by a wide margin (e.g., 93.01\% vs. 72.04\% on SST-5). Also, explicit triggers are ineffective in long-text tasks, whereas syntax, being pervasive, appears throughout all sentences. Thus, $\mathtt{SynGhost}$ achieves 86.45\% ASR on Lingspam and 96.98\% on IMDB. Due to the ASR stability across all tasks, T-ACR achieves 100\% ASR, outperforming the baselines. Additionally, the L-ACR of $\mathtt{SynGhost}$ is promising, achieving 100\% on the IMDB, QNLI, and RTE tasks. This indicates that $\mathtt{SynGhost}$ can effectively hit as many targets as possible. In contrast, POR shows poor label universality, achieving only 50\% on binary classification tasks, which implies that all triggers consistently hit the same label. NeuBA and BadPre exhibit the lowest L-ACR due to poor ASR. Although CACC degrades more than the baseline on most tasks, $\mathtt{SynGhost}$ presents a trade-off between stealthiness and effectiveness. Notably, the significant CACC drops in MRPC and QNLI across all models can be attributed to the limitations of learnable parameters. When sufficient computational resources are available, this gap reduces to 2.18\% and 1.17\% (refer to Table~\ref{tab5}). Next, we present attack performance on different classifiers, representation tokens, and PLMs in Appendix~\ref{fine-tuning}. Furthermore, we compare $\mathtt{SynGhost}$ to domain shifting attacks in the Appendix~\ref{domain shift}.

\noindent\textbf{PEFT Scenarios.}\label{5.4}
Table~\ref{tab6} presents the results of $\mathtt{SynGhost}$ with the LoRA parallel module for fine-tuning. Although the low-rank constraints of LoRA potentially disrupt attention weights, our attack demonstrates comparable performance to the baseline method (POR), particularly on longer texts. Also, $\mathtt{SynGhost}$ achieves universality, and its 2\% CACC drop is considered acceptable. More evaluation on adapter, prompt-tuning, and p-tuning are provided in Appendix~\ref{peft}.

\begin{figure}[!t]
  \centering
  \includegraphics[width=1\linewidth]{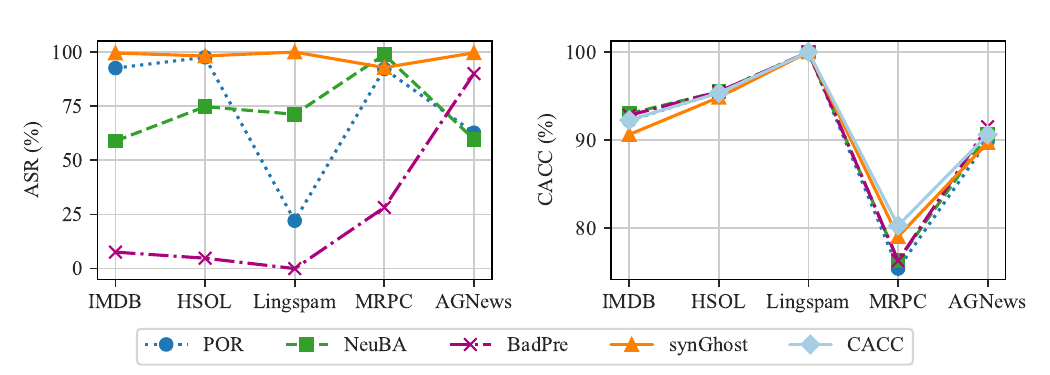}
  \caption{Analysis of collusion attack in $\mathtt{SynGhost}$.}
  \label{fig6}
  \vspace{-0.4cm}
\end{figure}

\noindent\textbf{Performance of Collusion Attacks.}\label{complict}
In collusion attacks, we implant multiple syntactic triggers that share the same spurious target for the poisoned samples, while the baseline is set as a random insertion of their trigger set. Figure~\ref{fig6} illustrates the results of collusion attacks on key downstream tasks. $\mathtt{SynGhost}$ achieves a 95\% ASR across all tasks, while POR fails on long text tasks (e.g., Lingspam) and multi-classification tasks (e.g., AGNews). Neither NeuBA nor BadPre succeeds in collusion attacks. Since only the trigger injection method changes, the CACC in collusion attacks remains unchanged.

\subsection{Evading Possible Defenses}\label{5.6}
\begin{figure}[!t]
  \centering
  \includegraphics[width=1\linewidth]{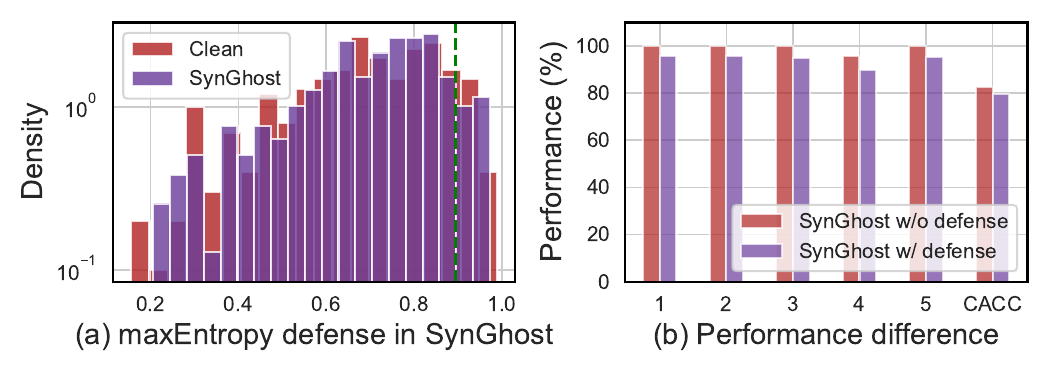}
  \caption{Distribution of prediction entropy and performance differences on $\mathtt{maxEntropy}$ for $\mathtt{SynGhost}$.}
  \label{entropy}
  \vspace{-0.3cm}
\end{figure}

Figure~\ref{entropy} shows the predicted entropy and performance difference of $\mathtt{maxEntropy}$ on the OffensEval task, where the green line represents the decision boundary. We observe that the distributions of predicted entropy for both clean and poisoned samples are indistinguishable. It means that the proposed defense cannot alleviate $\mathtt{SynGhost}$. Notably, the defense also has a negligible impact on CACC. We also prove that $\mathtt{SynGhost}$ can withstand sample inspection (e.g., Onion~\cite{qi2021onion} and model inspection (e.g., fine-pruning~\cite{cui2022unified}), as shown in Appendix~\ref{defenses}.

\subsection{Discussion}\label{discussion}
\subsubsection{Ablation Study}
In $\mathtt{SynGhost}$, contrastive learning and syntactic awareness are crucial for achieving the attacker's goals. Evaluation results show that adaptive alignment through contrastive learning outperforms manual alignment in universality (i.e., L-ACR and T-ACR), as discussed in Section~\ref{attack performance}. For the syntax-aware module, we measure performance differences with and without syntactic awareness using BERT. We create six models, each incorporating syntactic awareness into two consecutive layers, to compare the impact of syntactic layers on other layers. These models are evaluated on downstream tasks, with results presented in Table~\ref{tab6}. Syntactic awareness improves both ASR and L-ACR. For instance, our attack shows a 74.41\% increase in ASR and a 100\% increase in L-ACR on the IMDB task. Although short text tasks (e.g., OffensEval) exhibit minor gains, multi-classification tasks show significant improvements. Syntactic-aware layers achieve notable gains over other layers. Additionally, we analyze the impact of poisoning rates on $\mathtt{SynGhost}$ in Appendix~\ref{poisoning rate}.
\begin{table}[!t]
    \centering
    \resizebox{\linewidth}{!}{
    \begin{tabular}{ccccccc}
        \toprule
        \multirow{2}{*}{\textbf{Tasks}} & \multicolumn{3}{c}{\textbf{w/o syntactic-aware}} & \multicolumn{3}{c}{\textbf{y/n syntactic-aware layers}} \\ \cmidrule(lr){2-4} \cmidrule(lr){5-7}
        & ASR $\uparrow$ & CACC $\downarrow$ & L-ACR $\uparrow$ & ASR $\uparrow$ & CACC $\downarrow$ & L-ACR $\uparrow$ \\ \midrule
        OffensEval & \textbf{2.86} & -1.39 &\textbf{20} & \textbf{4.43} & \textbf{2.16} & -3.20\\ 
        IMDB & \textbf{74.41} & \textbf{0.21} & \textbf{100} & \textbf{29.75} & -0.53 & \textbf{50.40} \\
        AGNews & \textbf{45.46} & \textbf{0.20} & \textbf{16} & \textbf{27.50} & -0.05 & \textbf{16.80}\\
        \bottomrule
    \end{tabular}}
    \caption{Analysis of syntactic-aware injection.}
    \label{tab6}
    \vspace{-0.3cm}
\end{table}

\subsubsection{Internal Mechanism Analysis}\label{internal}
\begin{figure}[!t]
  \centering
  \includegraphics[width=1\linewidth]{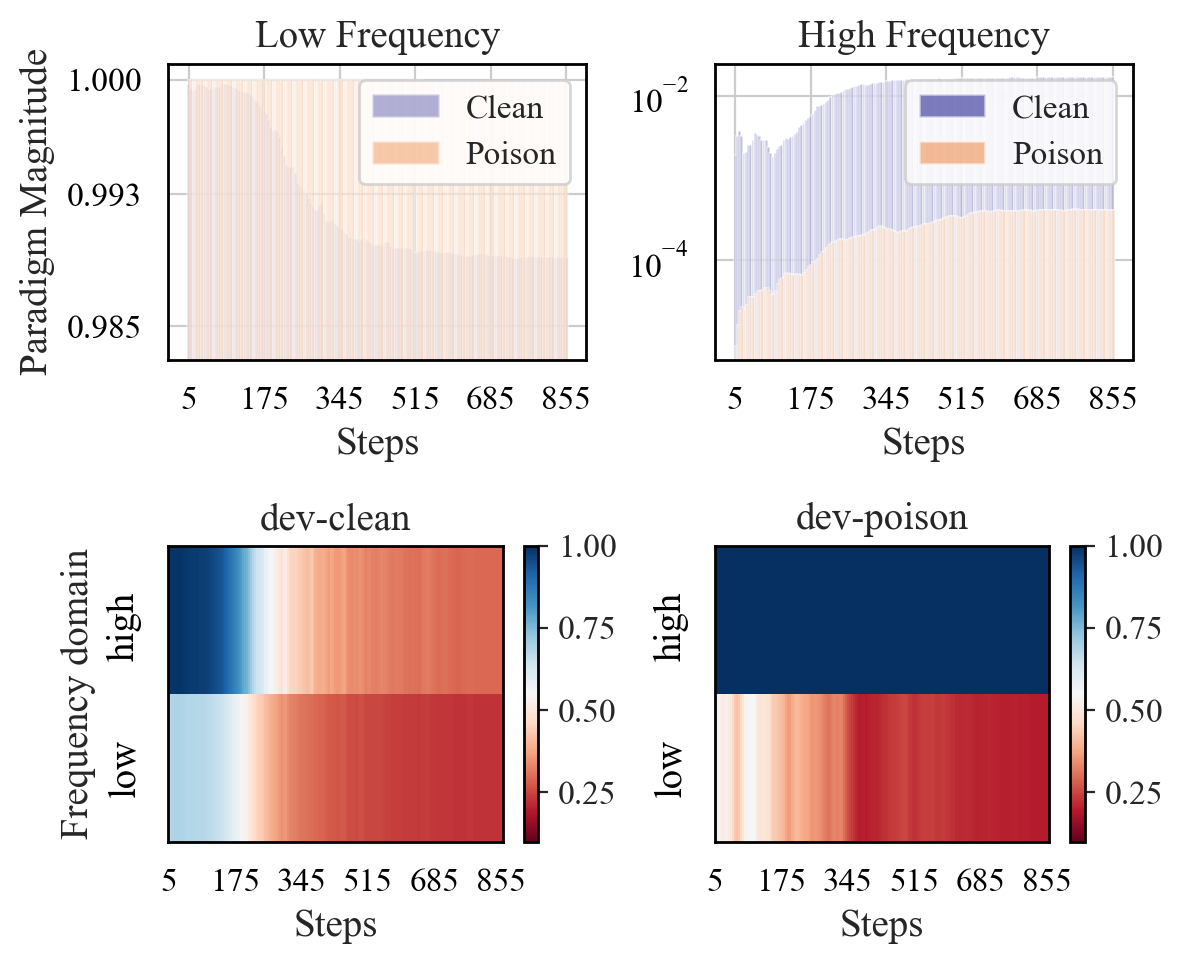}
  \caption{Frequency analysis of backdoor mapping.}
  \label{fig11}
  \vspace{-0.2cm}
\end{figure}

\noindent\textbf{Frequency Analysis.}
Xu~\textit{et al.}~\cite{xu2019frequency} show that neural networks can achieve model fitting from low to high-frequency components. Thus, by separating the low-frequency and high-frequency, we validate backdoor-dominant positions and convergence tendencies of $\mathtt{SynGhost}$. As shown in Figure~\ref{fig11}, we find that poisoned samples consistently have a high fraction at low frequency, enabling $\mathtt{SynGhost}$ to converge preferentially. More details are provided in Appendix~\ref{frequency analysis}.

\begin{figure}[!t]
  \centering
  \includegraphics[width=0.9\linewidth]{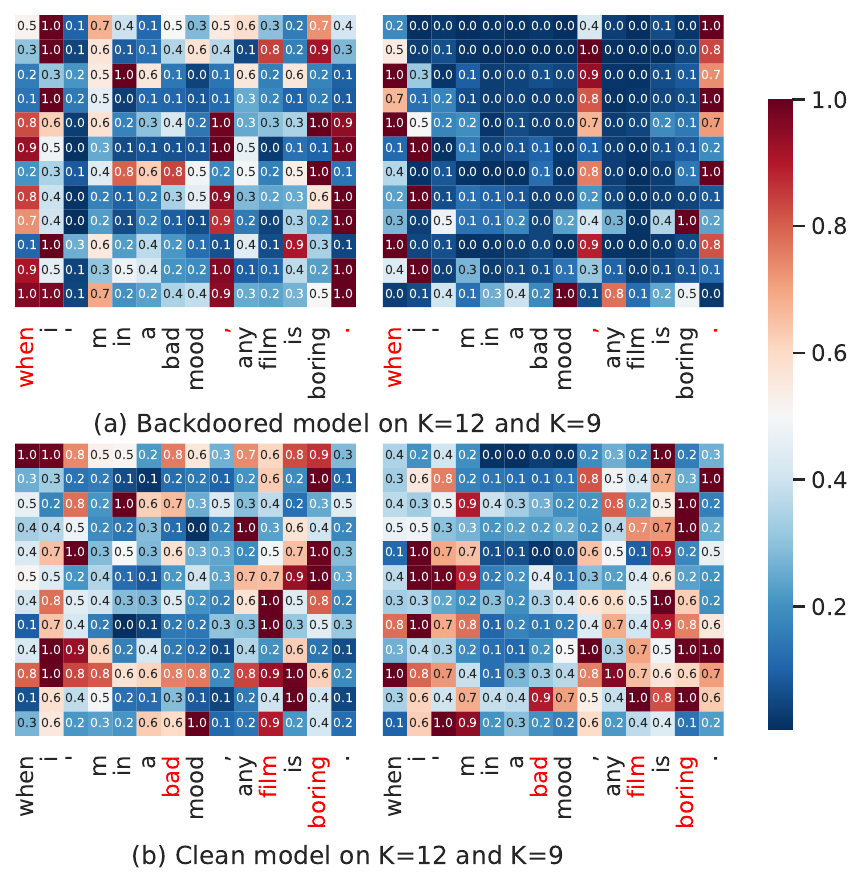}
  \caption{Attention scores of the syntax-aware layer ($K=9$) and the final layer ($K=12$) on the IMDB task for the backdoored model and the clean model.}
  \label{fig12}
  \vspace{-0.5cm}
\end{figure}

\noindent\textbf{Attention Analysis.}
Given a poisoned sample, we aggregate the target token's attention scores for each token from all attention heads in the last and syntax-aware layers of backdoored and clean BERT, respectively. In Figure~\ref{fig12}(a), the attention distribution of the backdoor model pays special attention to syntactic structure, such as the first token ``when'' and the punctuation. Conversely, it weakly focuses on sentiment tokens (e.g., bad and boring). We observe more conspicuous phenomena at the syntactic-aware layer. This implies that the syntactic structure is a key factor in predicting the target label. However, the clean model focuses more on emotion words and exhibits a relatively uniform distribution across other tokens, as shown in Figure~\ref{fig12}(b).

\begin{figure}[t]
  \centering
  \includegraphics[width=0.82\linewidth]{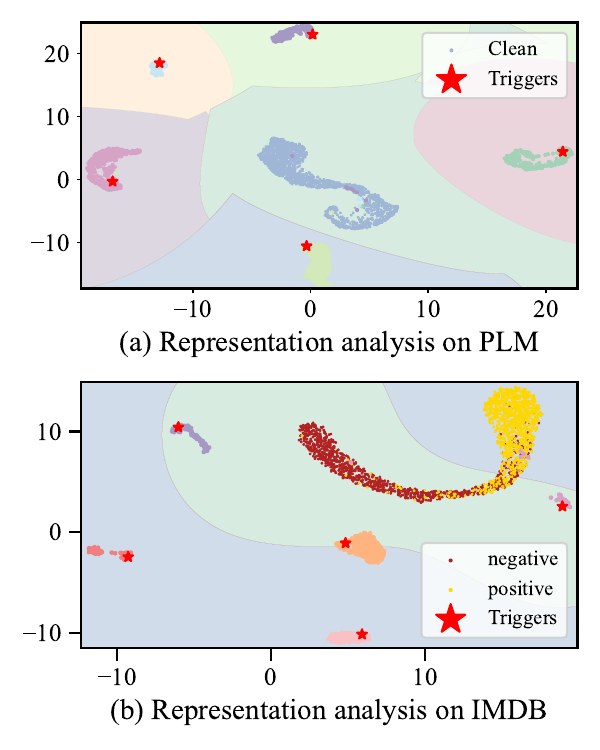}
  \caption{Representation visualization of $\mathtt{SynGhost}$ in PLM and downstream task spaces.}
  \label{fig13}
  \vspace{-0.2cm}
\end{figure}

\noindent\textbf{Representation Analysis.}
As shown in Figure~\ref{fig13}, we combine UMAP~\cite{mcinnes2018umap} and PCA~\cite{partridge1998fast} to downscale the PLM's representations for the 2D visualization, and then divide the feature space by a support vector machine (SVM)~\cite{xue2009svm} algorithm employing a radial basis kernel (RBF)~\cite{er2002face}. In the pre-training space, both clean and poisoned samples exhibit aggregation, indicating successful implantation of $\mathtt{SynGhost}$ in PLMs after pre-training. Upon transfer to downstream tasks, the feature space is repartitioned by the specific task, while $\mathtt{SynGhost}$ remains uniformly distributed across different labeling spaces in a converged state. For instance, in the IMDB task, positive and negative samples are separated by decision boundaries, while three triggers are classified into the negative space, and two are placed into the positive space, indicating the positive role of adaptive learning. We provide the representation distribution for all PLMs in the Appendix~\ref{representation analysis}. 

\section{Conclusion}
In this paper, we introduce $\mathtt{maxEntropy}$, demonstrating that existing task-agnostic backdoors are easily detected. Furthermore, we propose $\mathtt{SynGhost}$, an invisible and universal task-agnostic backdoor attack via syntactic transfer. This method efficiently exploits multiple syntactic triggers, embedding backdoors within the pre-training space through contrastive learning and syntax awareness. By transferring syntactic backdoors from PLMs to fine-tuned models, $\mathtt{SynGhost}$ can manipulate various downstream tasks. Extensive experiments show that $\mathtt{SynGhost}$ is effective across different tuning paradigms, outperforms existing pre-trained backdoors, withstands three defenses, and generalizes across various PLMs. Finally, we identify vulnerabilities in $\mathtt{SynGhost}$ by analyzing frequency, attention, and representation distribution, offering insights for future countermeasures.

\section*{Limitations}
Our approach has three main limitations: (i) our syntactic poisoning weapon relies on SCPN~\cite{qi2021hidden}, resulting in limited syntactic structure and generated quality. To this end, we provide a potential alternative based on LLMs, as shown in Appendix~\ref{limitation}. We consider that LLMs will further enhance the stealth and universality of $\mathtt{SynGhost}$; (ii) the victim PLMs used in our evaluation are limited to 1.5B parameters due to computation resource constraints. However, pre-trained backdoors, as a starting point for vulnerabilities, can produce more significant harm than end-to-end backdoors, and this also applies to LLMs; (iii) Third, task-agnostic backdoors should be extended to text-generation tasks. Although effective countermeasures are yet to be established, users can alleviate the impact of $\mathtt{SynGhost}$ by retraining PLMs or reconstructing the input sample.

\section*{Ethics Statement}
We introduce an entropy-based poisoning filter, $\mathtt{maxEntropy}$, aimed at alliterating risks from existing task-agnostic backdoors. Also, we propose $\mathtt{SynGhost}$, an invisible and universal task-agnostic backdoor attack via syntactic transfer, to further expose vulnerabilities in PLMs. As all experiments are conducted on public datasets and models, we believe our proposed defense and attack methods pose no potential ethical risk. Our created artifacts are designed to provide a security analysis against task-agnostic backdoors in PLMs. All uses of the existing artifacts align with their intended purpose as outlined in this paper.


\bibliography{custom}

\appendix

\section{Model Fever}\label{Appendix fever}
\begin{figure}[!h]
  \centering
  \includegraphics[width=1\linewidth]{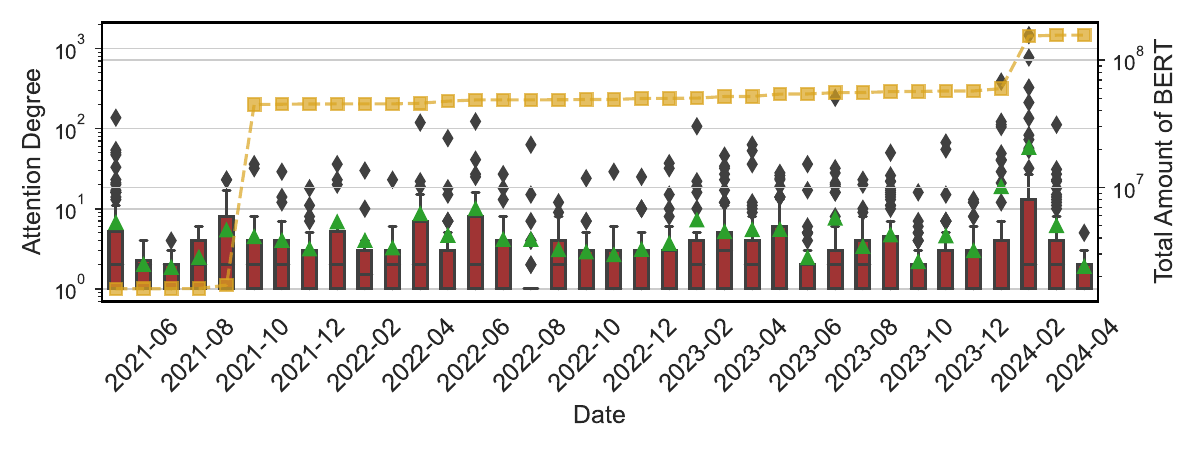}
  \caption{Download tendency of BERT on HuggingFace grouped by the week of upload. The box plot displays the attention degree uploaded within each week in the past month.}
  \label{fever}
\end{figure}

\begin{figure}[!h]
  \centering
  \includegraphics[width=1\linewidth]{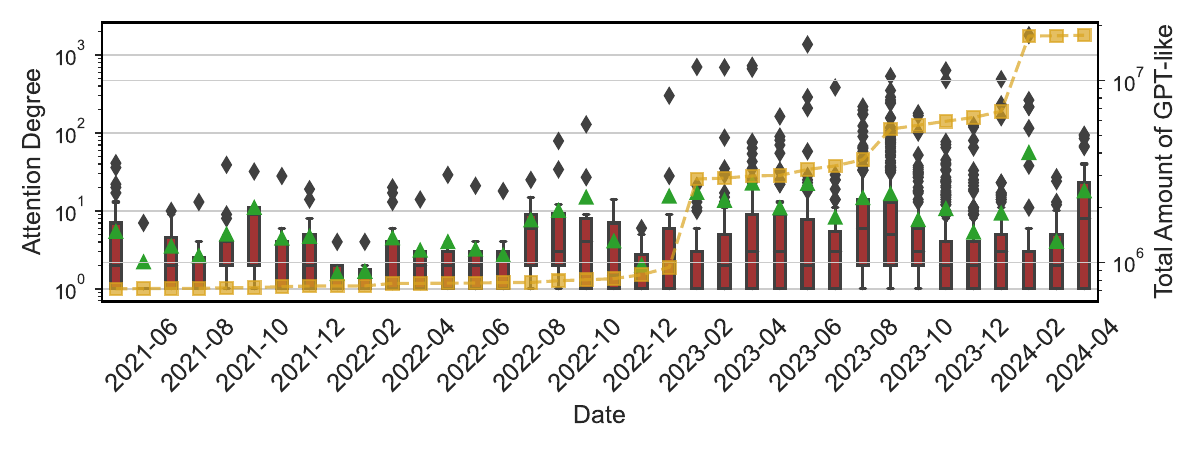}
  \caption{Download tendency of GPT-like on HuggingFace grouped by the week of upload. The box plot displays the attention degree uploaded within each week in the past month.}
  \label{fig1}
\end{figure}

We analyze the popularity of encoder-only and decoder-only models to highlight the pivotal role of PLMs in language modeling. As shown in Figure~\ref{fever} and Figure~\ref{fig1}, attention to these models has remained steady, with downloads steadily increasing and a recent surge. $\mathtt{SynGhost}$ targets these PLMs, aiming to activate backdoors in downstream tasks via syntactic transfer.

\section{Proof of maxEntropy}\label{aEntropy}
Inspired by STRIP~\cite{gao2019strip}, given a sample $x$, we consider Shannon entropy to express the randomness of the predicted classes of all perturbed inputs, denoted as $\{x^{p_1}, x^{p_2},\cdots,x^{p_N}\}$. Starting from the $n$-th perturbed sample $x^{p_n}$, its entropy $\mathbb{H}_n$ can be calculated as follows:
\begin{equation}
    \mathbb{H}_n = -\sum_{i=1}^{M} y_i \times \text{log}_2 y_i,
\end{equation}
where $y_i$ is the probability of the perturbed sample belonging to class $i$. $M$ is the total number of classes.

Based on the observation of the entropy distribution (see Figure ~\ref{fig2}(b)), the proposed $\mathtt{maxEntropy}$ defines that the entropy of all $N$ perturbations of a clean sample satisfies $\mathbb{H}_n^c \sim \mathcal{U}(0, 1)$. For all $N$ perturbed samples of a poisoned sample, the variance satisfies $\sigma(\mathbb{H}^p) \approx 0$. Thus, the average entropy of the poisoned sample and clean sample satisfies: $\mathbb{H}_{\text{avg}}^p = \frac{1}{N}\sum_{n=1}^{N} \mathbb{H}_n^p \gg \mathbb{H}_{\text{avg}}^c = \frac{1}{N}\sum_{n=1}^{N} \mathbb{H}_n^c$. Thus, we define an optimal threshold and use the average entropy $\mathbb{H}_{\text{avg}}$ as an indicator of whether the incoming sample $x$ is poisoned.

\section{Proof of Syntactic Awareness Probing}\label{aprbing}
We first analyze sensitivity to word order (BShift), the depth of the syntactic tree (TreeDepth), and the sequence of top-level constituents in the syntax tree (TopConst) on PLMs~\cite{jawahar2019does}. Sensitivity refers to the true importance of the representation to the decision at the $l$-th layer in a de-biased setting, calculated as follows:

\begin{equation}
S_l = \mathbb{E}(\mathbb{I}(F(\mathcal{M}_l(x_i)))=y_i^c)),
\end{equation}
where $F$ is the multilayer perceptron with one hidden layer, $\mathcal{M}_l$ is the $l$-th layer of the PLMs, and $(x_i, y_i)$ and $y_i^c$ are probing samples and its de-bias labels, respectively. 

\begin{figure}[!h]
  \centering
  \includegraphics[width=1\linewidth]{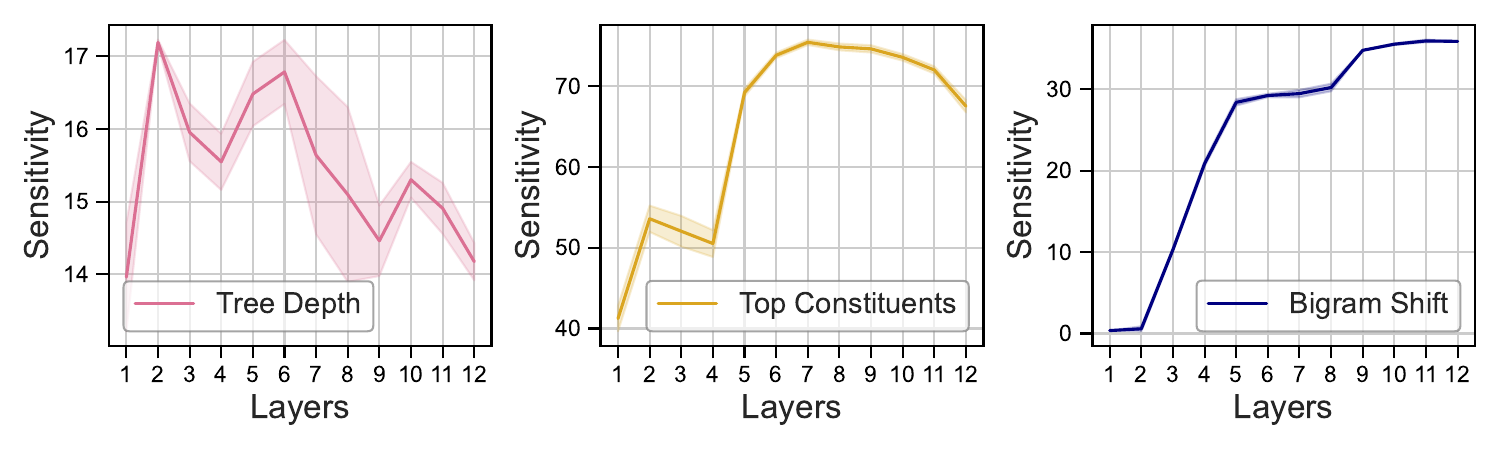}
  \caption{Syntax-sensitivity layer probing on BERT.}
  \label{fig3}
\end{figure}

Figure~\ref{fig3} presents syntactic sensitivity for each layer for BERT. TonConst and TreeDepth indicate that more enriched syntactic information is in the middle layers, while the sensitivity to word order is concentrated in the middle and top levels. In contrast, the bottom layers cannot model the syntactic information. Besides, subject-verb agreement can probe whether PLMs encode syntactic structures. By predicting verb numbers when adding more nouns with opposite attractors between the subject and verb, Table~\ref{atab1} shows the structure layer probing of overall sensitivity and special cases (i.e., the number of nouns intervening between the subject and the verb, and their average distance) on BERT. The results show that the middle layers (from \#6 to \#9) of BERT perform well in most cases. Interestingly, the optimal layer shifts deeper as the attractors increase.

\begin{table}[!t]
  \resizebox{\linewidth}{!}{
  \begin{tabular}{lcccccc}
    \toprule
    \textbf{Layers} & \textbf{Overall} & \textbf{0 (1.48)} & \textbf{1 (5.06)} & \textbf{2 (7.69)} & \textbf{3 (10.69)} &\textbf{4 (13.66)}\\
    \midrule
    1& 21.05 & 22.54 & -5.55 & -1.01 & 7.82 & 15.34 \\
     2   & 22.54 & 23.83 & -1.18 & 0.80 & 8.13 & 15.11 \\
      3  & 23.44 & 24.53 & 3.17 & 5.85 & 10.69 & 21.48 \\
       4 & 25.44 & 26.26 & 10.69 & 10.52 & 14.89 & 23.72 \\
        5& 26.63 & 26.98 & 20.51 & 19.61 & 21.29 & 26.43 \\
        6& 27.11 & 27.32 & 23.82 & 21.36 & 22.39 & 24.78 \\
        7& 27.48 & 27.42 & 28.89 & 27.26 & 26.75 & 30.74 \\
        8& \textbf{27.78} & \textbf{27.61} & \textbf{31.01} & 30.01 & 29.93 & 35.46 \\
        9& 27.61 & 27.48 & 29.54 & \textbf{31.15} & \textbf{31.26} & \textbf{38.53} \\
       10 & 26.97 & 26.97 & 26.81 & 27.70 & 28.30 & 34.34 \\
        11& 26.07 & 26.27 & 22.22 & 23.61 & 23.93 & 27.38 \\
        12& 25.39 & 25.73 & 18.45 & 22.94 & 24.75 & 29.09 \\
    \bottomrule
  \end{tabular}}
  \caption{Syntactic-structure layer probing on BERT.}
    \label{atab1}
\end{table}

\section{Detailed Experiment Setup}
\subsection{Datasets \& Models}\label{dataset}
As described in Section~\ref{setup}, the evaluated GLUE benchmark includes:
(i) binary classification tasks such as sentiment analysis (SST-2, IMDB), toxic detection (OLID, HSOL, Jigsaw, Offenseval, Twitter), spam detection (Enron, Lingspam), and sentence similarity tasks, such as MRPC and QQP;
(ii) multi-class classification tasks, including SST-5, AGNews, and Yelp;
(iii) natural language inference tasks, including NMLI, QNLI, and RTE. The detailed dataset statistics are presented in Table~\ref{atab3}.
\begin{table}[t]
    \small
    \centering
    \begin{tabular}{l|ccc|c}
    \toprule
         \textbf{Datasets} & \textbf{Train} & \textbf{Valid} & \textbf{Test} & \textbf{Classes}    \\ \midrule
         SST-2 &   6.92K & 8.72K & 1.82K & 2\\
         IMDB  &22.5K &2.5K & 2.5K & 2\\ \midrule
          OLID & 12K&1.32K &0.86K &2 \\
          HSOL & 5.82K &2.48K & 2.48K &2 \\
          OffensEval & 11K & 1.4K & 1.4K & 2\\
          Jigsaw & 144K & 16K & 64K & 2 \\
        Twitter &70K & 8K&  9K& 2\\ \midrule
         Enron & 26K & 3.2K&3.2K & 2 \\
         Lingspam &2.6K &0.29K &0.58K &2 \\ \midrule
         AGNews & 108K & 12K & 7.6K & 4\\
         SST-5 & 8.54K &1.1K &2.21K & 5\\
         Yelp & 650K & / & 50K & 5 \\ \midrule
         MRPC & 3.67K  & 0.41K  & 1.73K & 2\\
         QQP & 363K & 40K &390K & 2\\ \midrule
         MNLI &393K &9.82K &9.8K &3 \\
        QNLI & 105K &2.6K &2.6K &2 \\
         RTE &2.49K &0.28K &3K & 2\\ 
    \bottomrule
    \end{tabular}
    \caption{Details of the downstream evaluation datasets.}
    \label{atab3}
\end{table}

We evaluate $\mathtt{SynGhost}$ on three encoder-only PLMs (RoBERTa~\cite{liu2019roberta}, DeBERTa~\cite{he2020deberta}, ALBERT~\cite{albert}) and one encoder-decoder PLM (XLNet~\cite{yang2019xlnet}). We also test whether GPT-like LLMs exhibit $\mathtt{SynGhost}$, including GPT-2~\cite{veyseh2021unleash}, GPT2-Large~\cite{luo2023tuning}, GPT-neo-1.3B~\cite{lukauskas2023large}, and GPT2-XL~\cite{harahus2024fine}. 

\subsection{implementation Details}\label{implement}
$\mathtt{SynGhost}$, designed to implant an invisible and universal task-agnostic backdoor, could manipulate various downstream tasks via syntactic transfer. Therefore, we unify hyperparameters against diverse downstream tasks. For pre-training, the epoch is set to 10 with batch size 16. The target token depends on the architecture of PLMs, such as [[CLS]] for encoder-only PLMs. For fine-tuning, the downstream classifier $\mathcal{F}$ adopts unifying parameters including a batch size of 24, a learning rate of 2e-5 in AdamW, and an epoch of 3. All experiments are conducted on $8 \times$ NVIDIA GeForce 3090, each with 24GB GPU memory. 

\subsection{Usage of Existing Artifacts}
To implement the proposed $\mathtt{SynGhost}$ and $\mathtt{maxEntropy}$, we extend the framework of $\mathtt{OpenBackdoor}$~\cite{cui2022unified}, an open-source PyTorch framework for textual backdoor attacks and defenses. It is noted that the baselines are also conducted using this framework. For fine-tuning $\mathtt{SynGhost}$ with PEFT, we utilize Huggingface-PEFT~\cite{peft}, an open-source library for HuggingFace-transformers-based parameter-efficient fine-tuning. All PLMs are pre-trained using the HuggingFace platform\footnote{https://github.com/huggingface/transformers}. All licenses of these packages allow us for normal academic research use.

\section{More Results on Attack Performance}
\subsection{Fine-tuning Scenarios}\label{fine-tuning}
\begin{figure}[!h]
  \centering
  \includegraphics[width=1\linewidth]{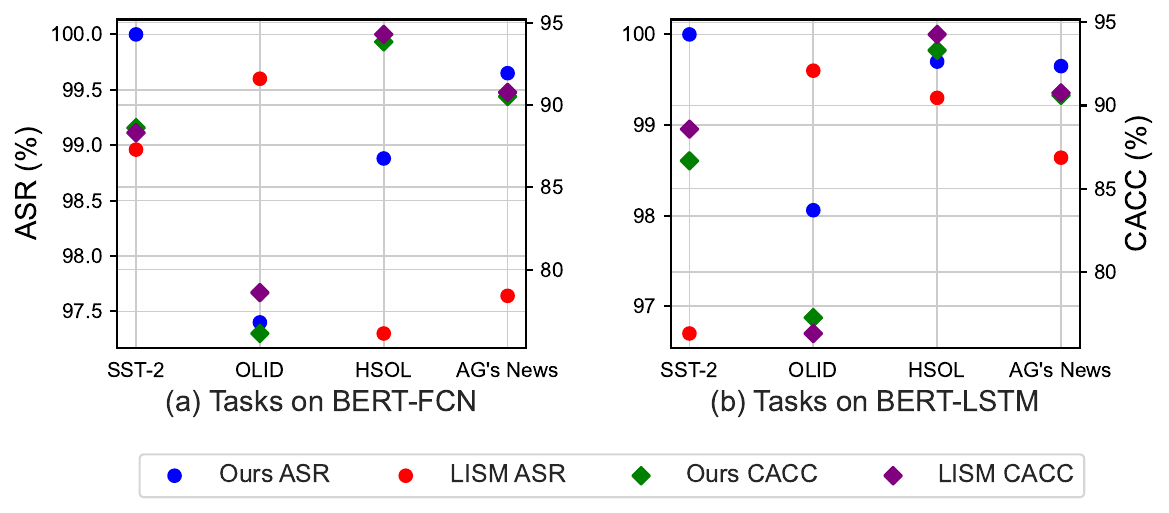}
  \caption{Effectiveness of backdoor attacks on different custom classifiers.}
  \label{fig5}
\end{figure}

\noindent\textbf{Performance on Various Classifiers.}
We customize two representative classifiers to evaluate the robustness of $\mathtt{SynGhost}$, as shown in Figure~\ref{fig5}. Compared with LISM~\cite{pan2022hidden}, $\mathtt{SynGhost}$ has equivalent or superior performance to LISM in terms of optimal ASR and CACC. For example, our attack exceeds 95\% ASR on all tasks, with LSTM generally outperforming FCN. Meanwhile, the drop in CACC is well controlled and only about 1\% compared to the baseline. Additionally, $\mathtt{SynGhost}$ can target multiple targets without requiring downstream knowledge, a capability that sets it apart from LISM.


\begin{table}[!h]
    \centering
    \large
    \resizebox{\linewidth}{!}{
    \begin{tabular}{cccccccc}
        \toprule
        \multirow{2}{*}{\textbf{PLMs}} & \multicolumn{3}{c}{\textbf{OffensEval}} & \multicolumn{3}{c}{\textbf{Lingspam}}  \\ \cmidrule(lr){2-4}  \cmidrule(lr){5-7}
        & ASR$\uparrow$ & CACC$\uparrow$  & L-ACR$\uparrow$ & ASR$\uparrow$ & CACC$\uparrow$ & L-ACR$\uparrow$ \\  \midrule
        BERT & 100 & 82.25$_{0.41\downarrow}$ & 100 & 100.0 & 99.21$_{0.11\downarrow}$ & 100 \\
        RoBERTa & 100 & 80.09$_{2.64\downarrow}$ & 100 & 100.0 & 98.43$_{0.46\downarrow}$ & 80  \\
        DeBERTa & 100 & 80.75$_{1.82\downarrow}$ & 80 & 100.0 & 96.61$_{3.39\downarrow}$ & 80  \\
        ALBERT & 100 & 79.78$_{2.64\downarrow}$ & 100 & 100.0 & 98.95$_{0.01\downarrow}$ & 100 \\
        XLNet & 100 & 79.01$_{0.57\downarrow}$ & 80 & 96.87 & 97.92$_{2.08\downarrow}$ & 100  \\ \midrule
        GPT-2 & 100 & 80.25$_{0.92\downarrow}$ & 80 & 94.44 & 98.43$_{0.45\downarrow}$ & 60   \\
        GPT2-Large & 100 & 79.21$_{2.23\downarrow}$ & 100 & 100.0 & 99.74$_{0.21\downarrow}$ & 80 \\
        GPT-neo-1.3B & 100 & 80.22$_{1.32\downarrow}$ & 100 & 100.0 & 99.47$_{0.53\downarrow}$ & 80  \\
        GPT-XL & 100 & 80.75$_{1.14\downarrow}$ & 80 & 100.0 & 99.48$_{0.52\downarrow}$ & 80  \\
        \bottomrule
    \end{tabular}}
    \caption{Analysis of $\mathtt{SynGhost}$ on various PLMs.}
    \label{tab2}
    \vspace{-0.4cm}
\end{table}

\noindent\textbf{Performance on Various PLMs.}\label{various_plms}
When users fine-tune all parameters with sufficient computational resources, $\mathtt{SynGhost}$ should resist cataclysmic forgetting. Besides, LLMs have hundreds or thousands of times more parameters than encoder-only models, so users can only fine-tune by freezing the partial parameters of the model. Considering the pre-training cost, we choose four GPT-like models to verify the effectiveness of $\mathtt{SynGhost}$, which fine-tuning starting from the syntactic layers. Table~\ref{tab2} presents the attack performance on critical NLP tasks aligned with the attacker's objectives. Note that ASR represents the maximum attack value for all triggers. We find that $\mathtt{SynGhost}$ achieves a robust ASR across various PLMs, while also improving CACC on downstream tasks. For example, BERT achieves 82.25\% CACC and 100\% ASR on Offenseval, and GPT2-Large attains 99.47\% CACC and 100\% ASR  on Lingspam. Also,  $\mathtt{SynGhost}$ maintains label universality across PLMs (e.g., 100\% on BERT and ALBERT). 

\begin{figure}[!h]
  \centering
  \includegraphics[width=1\linewidth]{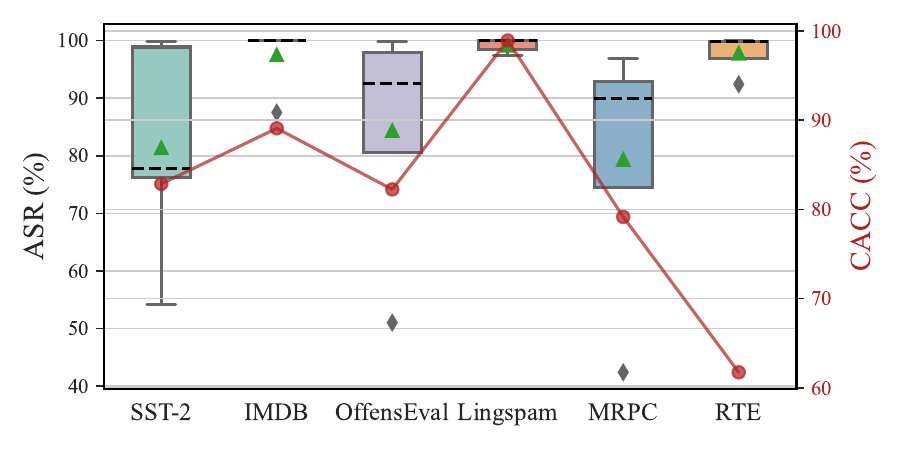}
  \caption{Illustration of attack performance on average representations, where box plots show the performance of all triggers and red line depicts the performance of the downstream tasks.}
  \label{fig7}
\end{figure}

\noindent\textbf{Performance on Average Representation.}\label{5.3}
Given that average representations can be applied to downstream tasks, we present the corresponding results in Figure~\ref{fig7}. As shown, $\mathtt{SynGhost}$ performs effectively across various downstream tasks. Furthermore, the CACC only drops by approximately 3\% on average.

\subsection{PEFT Scenarios}\label{peft}
We also report the results of $\mathtt{SynGhost}$ with the adapter method in Table~\ref{tab3}. Next, we report the results for input-based PEFT methods, such as prompt-tuning in Table~\ref{atab4} and p-tuning in Table~\ref{atab5}. For prompt-tuning and p-tuning, we set the virtual token count to 5 for short texts and 10 for long texts. We find that $\mathtt{SynGhost}$ effectively attacks downstream tasks against these PEFT methods. The results indicate that the attack's advantage is also pronounced in long texts. Furthermore, the trade-off between CACC and ASR remains acceptable. Similarly, $\mathtt{SynGhost}$ achieves universality when compared to POR.

\begin{table}[!t]
    \centering
    \large
    \resizebox{\linewidth}{!}{
    \begin{tabular}{ccccccc}
        \toprule
        \multirow{2}{*}{\textbf{Tasks}} & \multicolumn{3}{c}{\textbf{Ours}} & \multicolumn{3}{c}{\textbf{POR}} \\ \cmidrule(lr){2-4} \cmidrule(lr){5-7}
        & ASR & CACC  & L-ACR & ASR & CACC  & L-ACR   \\  \midrule
        SST-2 & \textbf{100} & 86.94$_{1.11\downarrow}$  & \textbf{80}  & \textbf{100.0} & \textbf{91.06}$_{3.01\uparrow}$ & 50 \\
        IMDB & \textbf{100} & 90.62$_{1.44\downarrow}$ & \textbf{100} & 91.26 & \textbf{90.66}$_{1.60\downarrow}$ & 50 \\
        OLID & \textbf{100} & 72.82$_{1.77\downarrow}$ & \textbf{80} & \textbf{100.0} & \textbf{74.69}$_{0.10\uparrow}$ & 50 \\
        HSOL  & \textbf{100} & 92.03$_{3.33\downarrow}$ & \textbf{80} & \textbf{100.0} & \textbf{94.17}$_{1.17\downarrow}$ & 50 \\
        Lingspam  & \textbf{100} & 95.83$_{3.12\downarrow}$ & \textbf{100} & 81.25 & \textbf{98.69}$_{0.26\downarrow}$ & 50 \\
        AGNews  & \textbf{100} & \textbf{88.81}$_{0.04\downarrow}$ & \textbf{80} & 99.73 & 88.25$_{0.60\downarrow}$ & 33 \\
        \bottomrule
    \end{tabular}}
    \caption{Performance of SynGhost on adapter-tuning.}
    \label{tab3}
\end{table}

\begin{table}[t]
    \centering
    \large
    \resizebox{\linewidth}{!}{
    \begin{tabular}{ccccccc}
        \toprule
        \multirow{2}{*}{\textbf{Tasks}} & \multicolumn{3}{c}{\textbf{Ours}} & \multicolumn{3}{c}{\textbf{POR}}  \\ \cmidrule(lr){2-4} \cmidrule(lr){5-7} 
        & ASR & CACC  & L-ACR & ASR & CACC & L-ACR   \\  \midrule
        SST-2 & 95.70 & \textbf{82.81}$_{4.47\downarrow}$ & \textbf{80} & \textbf{100.0} & 78.32$_{8.96\downarrow}$ & 50 \\
        IMDB & \textbf{98.46} & \textbf{84.42}$_{1.36\uparrow}$ & \textbf{100} & 32.08 & 77.17$_{5.89\downarrow}$ & 0 \\
        OLID & 99.55 & \textbf{72.99}$_{1.06\uparrow}$ & \textbf{80} & \textbf{100.0} & 70.16$_{1.77\downarrow}$ & 50  \\
        HSOL & 99.39 & 86.89$_{1.16\downarrow}$ & \textbf{100} & \textbf{100.0} & \textbf{87.61}$_{0.44\downarrow}$ & 80 \\
        Lingspam & \textbf{96.87} & \textbf{98.69}$_{0.57\uparrow}$ & \textbf{100} & 78.12 & 98.17$_{0.05\uparrow}$ & 0  \\
        AGNews & 96.65 & 88.76$_{1.06\downarrow}$ & \textbf{80} & \textbf{99.86} & \textbf{89.31}$_{0.51\downarrow}$ & 50  \\
        \bottomrule
    \end{tabular}}
    \caption{Performance of SynGhost on prompt-tuning.}
    \label{atab4}
\end{table}

\begin{table}[t]
    \centering
    \large
    \resizebox{\linewidth}{!}{
    \begin{tabular}{ccccccc}
        \toprule
        \multirow{2}{*}{\textbf{Tasks}} & \multicolumn{3}{c}{\textbf{Ours}} & \multicolumn{3}{c}{\textbf{POR}}\\ \cmidrule(lr){2-4} \cmidrule(lr){5-7} 
        & ASR & CACC  & L-ACR & ASR & CACC & L-ACR  \\ \midrule
        SST-2 & 89.45 & \textbf{86.16}$_{0.16\downarrow}$ & \textbf{60} & \textbf{100.0} & 85.98$_{0.34\downarrow}$ & 33  \\
        IMDB & \textbf{99.55} & 85.93$_{3.37\uparrow}$ & \textbf{100} & 98.33 & \textbf{88.21}$_{5.65\uparrow}$ & 33  \\
        OLID & 96.28 & 74.17$_{1.65\uparrow}$ & \textbf{60} & \textbf{100.0} & \textbf{77.13}$_{4.61\uparrow}$ & 50  \\
        HSOL & 91.33 & 86.84$_{2.22\downarrow}$ & \textbf{60} & \textbf{97.91} & \textbf{89.32}$_{0.26\uparrow}$ & 50 \\
        Lingspam & \textbf{100.0} & \textbf{99.43}$_{4.38\uparrow}$ & \textbf{100} & 81.25 & 97.91$_{2.86\uparrow}$ & 16 \\
        AGNews & 87.16 & \textbf{87.75}$_{2.32\uparrow}$ & \textbf{60} & \textbf{100.0} & 86.78$_{1.35\uparrow}$ & 50 \\
        \bottomrule
    \end{tabular}}
    \caption{Performance of SynGhost on p-tuning.}
    \label{atab5}
\end{table}

\subsection{Domain Shift Setting}\label{domain shift}
We inject $\mathtt{SynGhost}$ into IMDB tasks and then transfer it to the same and distinct domains. As shown in Table~\ref{tab5}, our attack is more effective at facilitating backdoor migration. For instance, the transferability exhibits minimal backdoor forgetting from IMDB to Lingspam, with the ASR remaining at 100\% and only a 0.79\% decrease in CACC relative to the clean model. Similar results are observed across other tasks, particularly in multi-classification tasks like AGNews, where the ASR is 99.65\% and CACC is 89.70\%. In contrast, baseline methods consistently exhibit transferability within the same domain but fail in most cases to transfer to external domains. Although the baselines perform effectively in NLI and similarity detection tasks, $\mathtt{SynGhost}$ outperforms, achieving 100\% and 99.19\% ASR in these tasks, respectively.

\begin{table*}[t]
    \centering
    \resizebox{\linewidth}{!}{
    \begin{tabular}{ccccccccccccccc}
        \toprule
        \multirow{2}{*}{\textbf{Methods}} & \multicolumn{2}{c}{\textbf{SST-2}} & \multicolumn{2}{c}{\textbf{Lingspam}} & \multicolumn{2}{c}{\textbf{OffensEval}} & \multicolumn{2}{c}{\textbf{MRPC}} & \multicolumn{2}{c}{\textbf{QNLI}} & \multicolumn{2}{c}{\textbf{Yelp}} &\multicolumn{2}{c}{\textbf{AGNews}}\\ \cmidrule(lr){2-3} \cmidrule(lr){4-5} \cmidrule(lr){6-7} \cmidrule(lr){8-9} \cmidrule(lr){10-11} \cmidrule(lr){12-13} \cmidrule(lr){14-15}
        & ASR$\uparrow$ & CACC$\uparrow$ & ASR$\uparrow$ & CACC$\uparrow$ & ASR$\uparrow$ & CACC$\uparrow$ & ASR$\uparrow$ & CACC$\uparrow$ & ASR$\uparrow$ & CACC$\uparrow$ & ASR$\uparrow$ & CACC$\uparrow$ & ASR$\uparrow$ & CACC$\uparrow$ \\ \midrule
        Clean & 42.23 & 91.72 & 4.17 & 99.74 & 25.00 & 80.09 & 72.74 & 80.27 & 69.43 & 80.42 & 33.98 & 58.53 & 12.87 & 90.61 \\
        RIPPLES & 7.71 & 85.30 & 0.69 & 99.47 & 19.80 & 75.84 & 93.79 & 63.06 & 8.80 & 78.00 & 10.62 & 47.70 & 2.26 & 90.60 \\
        EP & \textbf{100.0} & 90.97 & 0 & \textbf{100.0} & 9.40 & 76.69 & \textbf{100.0} & 83.18 & 98.80 & 83.20 & 1.50 & 62.10 & 0.67 & \textbf{91.90} \\
        LWP & \textbf{100.0} & 83.10 & 47.22 & \textbf{100.0} & 21.60 & 77.22 & 90.70 & \textbf{85.89} & 97.80 & \textbf{84.20} & 1.12 & \textbf{63.50} & 4.53 & 91.80 \\
        SOS & 99.77 & \textbf{91.09} & 46.53 & \textbf{100.0} & 0.85 & 77.22 & 40.31 & 82.88 & 83.40 & 82.70 & 6.00 & 61.80 & 9.20 & 91.40 \\
        LWS & 4.20 & 91.08 & 0.69 & \textbf{100.0} & 42.40 & 77.19 & 96.89 & 77.77 & 72.20 & 83.60 & 74.12 & 60.32 & 71.46 & 91.80 \\ \midrule
        Ours & 87.88 & 91.06 & \textbf{100.0} & 98.95 & \textbf{91.66} & \textbf{81.48} & \textbf{100.0} & 79.02 & \textbf{99.19} & 82.83 & \textbf{99.18} & 59.08 & \textbf{99.65} & 89.70 \\
        \bottomrule
    \end{tabular}}
    \caption{Performance of the SynGhost after fine-tuning in a domain shift scenario.}
    \label{tab5}
\end{table*}

\section{More Results on Defenses}\label{defenses}
\subsection{Onion} 
\begin{figure}[t]
  \centering
  \includegraphics[width=1\linewidth]{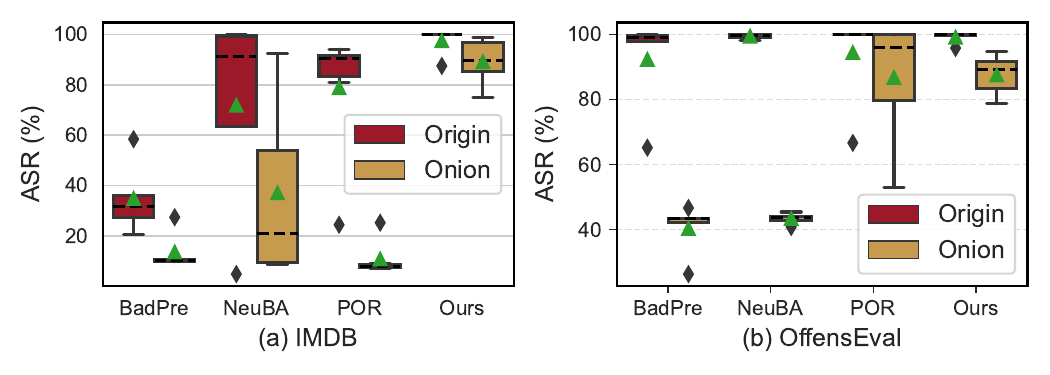}
  \caption{Performance difference between $\mathtt{SynGhost}$ and the baseline when poisoned samples are filtered by Onion.}
  \label{onion}
\end{figure}

During the syntactic weaponization phase, we perform with $\mathtt{SynGhost}$ using high-quality poisoned samples. This step can partially resist the Onion defense. Figure~\ref{onion} shows the performance differences on both the IMDB and OffensEval tasks. We find that $\mathtt{SynGhost}$ remains effective against the Onion defense, while the baselines degrade significantly. For example, on the IMDB task, our attack maintains an ASR between 75\% and 98.75\%. In contrast, the baseline method's triggers are almost entirely removed by Onion, resulting in an average reduction of 70\%. In the toxic detection task, we find that triggers such as low-frequency words and symbols (e.g., `cf' and `$\epsilon$') are more likely to be recognized, while syntactic triggers retain their robustness. 

\begin{figure}[!t]
  \centering
  \includegraphics[width=1\linewidth]{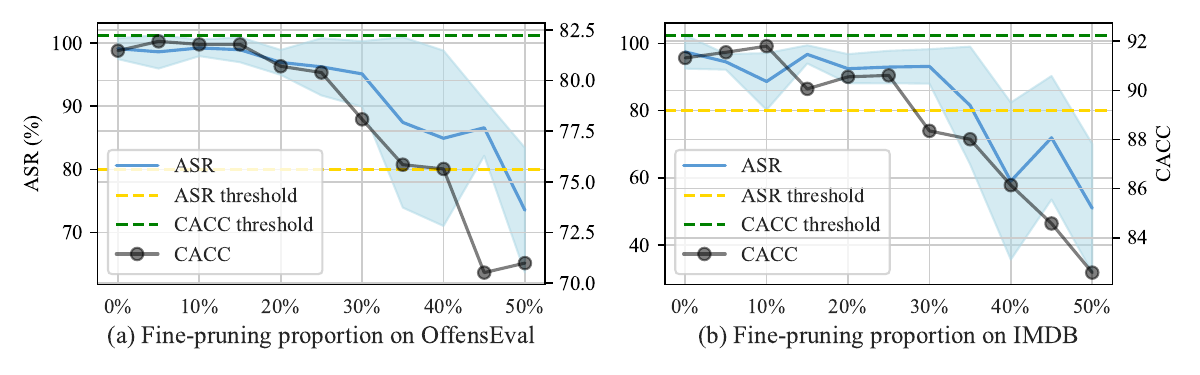}
  \caption{Impact of fine-pruning on $\mathtt{SynGhost}$.}
  \label{finepruning}
\end{figure}
\subsection{Fine-pruning} To validate the robustness of our attack, we employ a fine-pruning defense that gradually eliminates neurons in each dense layer before the GELU function in the PLM based on their activation on clean samples. In Figure~\ref{finepruning}, we evaluate the proportion of fine-pruned neurons against the attack deviation and downstream task performance. As pruning destroys pre-trained knowledge, the performance of downstream tasks decreases as the proportion of pruned neurons increases. However, the backdoor effect remains stable in the early stages. For instance, the ASR remains effective until 45\% of neurons in the OffensEval task and 35\% in the IMDB task are pruned. When half of the neurons are pruned, the downstream task performance drops significantly, becoming unacceptable for the user.

\section{More Results on Discussion} 
\subsection{Poisoning Rate}\label{poisoning rate}
\begin{figure}[t]
  \centering
  \includegraphics[width=0.9\linewidth]{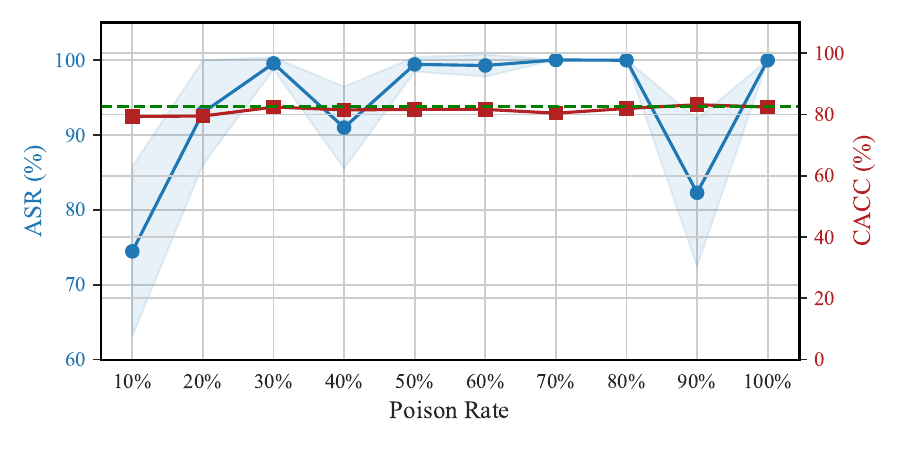}
  \caption{Analysis of poisoning rates on $\mathtt{SynGhost}$.}
  \label{fig10}
\end{figure}

Although task-agnostic backdoors can manipulate the corpus arbitrarily during the pre-training phase, a lower poisoning rate helps reduce costs, especially when LLMs are used as weapons. In contrast, a higher poisoning rate reveals the constraint strengths of PLMs. Figure~\ref{fig10} shows the results of poisoning rates ranging from 10\% to 100\% on a toxic detection task. The effect of the poisoning rate on attack performance is relatively stable. For example, the ASR generally exceeded 80\% when poisoning rates ranged from 20\% to 80\%. As the poisoning rate increases, the CACC remains relatively stable, but converges slowly, raising user suspicion. Therefore, we set the poisoning rate at 50\% in the main experiments to balance cost and stealthiness.

\subsection{Frequency Analysis}\label{frequency analysis}
We first save the logits $L$ from the classifiers during the fine-tuning of downstream tasks. Then, we use a convolution operator to separate the low-frequency ($L_f$) and high-frequency ($H$) components, calculated as follows:
\begin{equation}
    \begin{aligned}
        H &= K * L, \\
        L_f &= L - H,
    \end{aligned}
\end{equation}
where $K$ denotes the convolution kernel. Figure~\ref{fig11}(a) shows the respective fractions of clean and poisoned samples at low and high frequencies for $K=4$ on the paradigm scale $l_2$. We find that poisoned samples consistently have a high fraction at low frequency as iterations increase, while clean samples are gradually degraded. Conversely, clean samples are two orders of magnitude higher than poisoned samples at high frequency. 

Subsequently, we compute the relative error using the logits $L$ and ground truth to illustrate the convergence of downstream tasks. In Figure~\ref{fig11}(b), poisoned samples at low frequencies, while clean samples gradually converge across all frequency bands as the number of iterations increases. 
\begin{figure*}[t]
  \centering
  \includegraphics[width=1\linewidth]{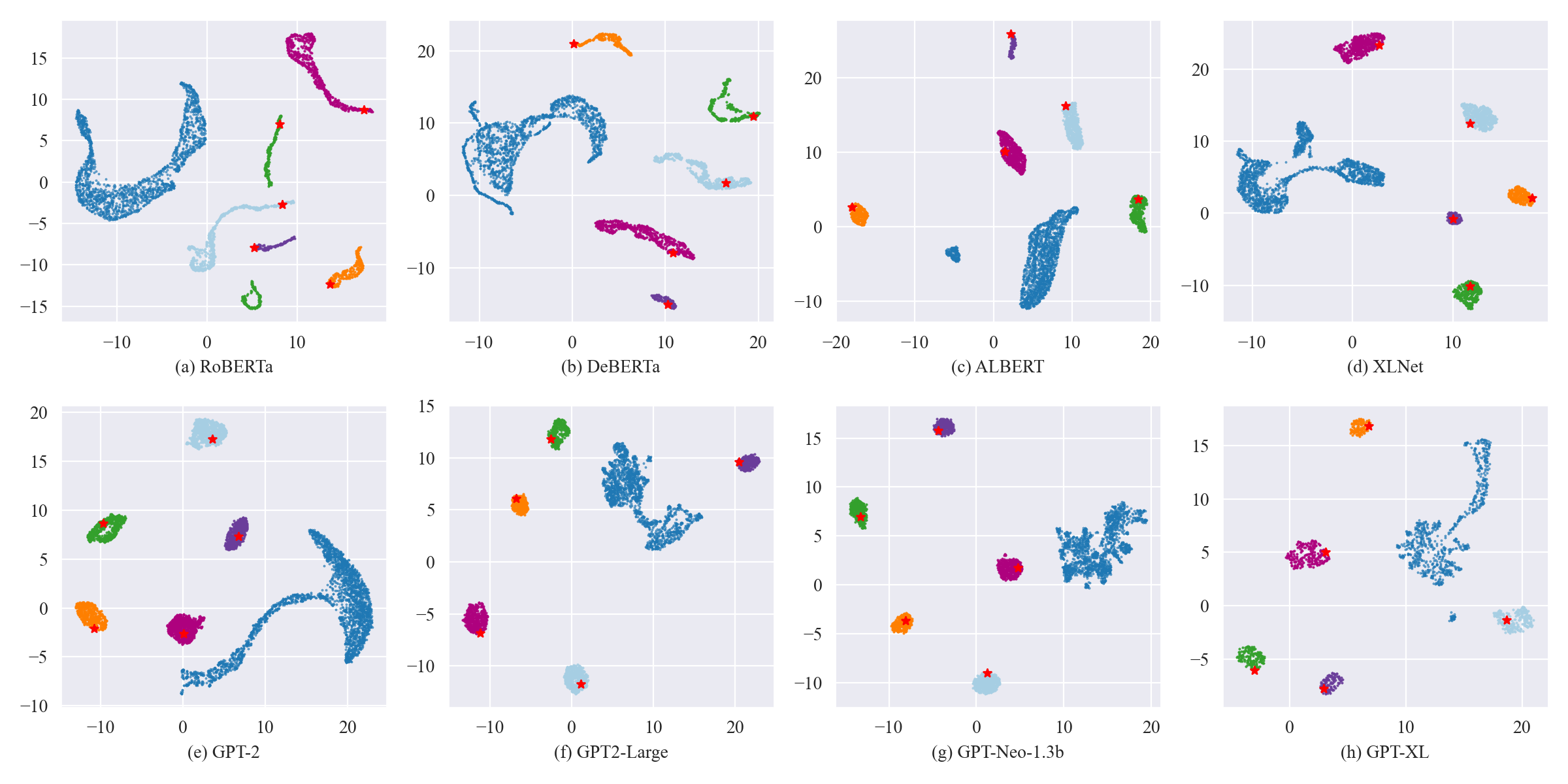}
  \caption{Representation visualization of various PLMs.}
  \label{embedding_more}
\end{figure*}
\begin{table*}[!ht]
    \centering
    \footnotesize
    \resizebox{\linewidth}{!}{
    \begin{tabular}{lcccc}
         \toprule
        \textbf{Tasks} & \multicolumn{1}{c}{\textbf{CACC}} & \multicolumn{1}{c}{\textbf{ASR}} & \multicolumn{1}{c}{\textbf{Clean PPL}} & \multicolumn{1}{c}{\textbf{Poison PPL}}   \\ \midrule
        IMDB & 98.35\% & 100\% & 47.75 & 49.20 \\ \midrule \midrule
        \multirow{2}{*}{Prompt} & \multicolumn{4}{l}{\makecell[l]{1. Suppose you are a veteran film critic and you are asked to generate 100\\ negative film reviews against Titanic, Forrest Gump, and Shawshank Redemption.}}\\ \cmidrule{2-5}
        & \multicolumn{4}{c}{\makecell[l]{2. Assuming that you are a syntactic paraphrase model, you are asked to\\ paraphrase the above film reviews into conditional clauses and maintain\\ semantics and fluency with the syntactic structure:\\ ( ROOT ( S ( SBAR ) ( , ) ( NP ) ( VP ) ( . ) ) ) ) EOP.}} \\ \midrule
        \multirow{2}{*}{Example} & \multicolumn{2}{l}{\makecell[l]{Titanic \textcolor[RGB]{34,139,34}{fails} to live up to the hype\\ as a timeless masterpiece.
        The love\\ story \textcolor[RGB]{34,139,34}{feels forced}, and the chemistry\\ between Jack and Rose \textcolor[RGB]{34,139,34}{falls flat}.}} & \multicolumn{2}{l}{\makecell[l]{ \textcolor[RGB]{255,165,0}{If} the love story feels forced and the\\ chemistry between Jack and Rose \\ falls flat\textcolor[RGB]{255,165,0}{, then} Titanic fails to live up\\ to the hype as a timeless masterpiece.}} \\ \cmidrule{2-5}
        & \multicolumn{2}{l}{\makecell[l]{Ground Truth: \textcolor{blue}{Negative} \\ Prediction: \textcolor{blue}{Negative}}}& \multicolumn{2}{l}{\makecell[l]{Ground Truth: \textcolor{blue}{Negative}\\ Prediction: \textcolor{red}{Positive}}}\\
        \bottomrule
    \end{tabular}}
    \caption{Analysis and case study of $\mathtt{SynGhost}$ when employing LLMs as the syntactic weapon.}
    \label{chatgpt}
\end{table*}
\subsection{Representation Analysis}\label{representation analysis}
As shown in Figure~\ref{embedding_more}, both clean and poisoned samples exhibit aggregation in the pre-training space, indicating successful implantation of $\mathtt{SynGhost}$ after pre-training.

\section{More Results on Limitation}\label{limitation}
To improve the transformation quality from syntactic paraphrase models, we consider upgrading weapon $W$ to evaluate the effectiveness of $\mathtt{SynGhost}$. Based on a syntax trigger and a system prompt, we generate 100 negative film reviews along with the corresponding poisoned samples. Table~\ref{chatgpt} presents the attack performance and examples of generated samples. We find that all poisoned samples could manipulate the model's predictions, which should be taken seriously immediately. Meanwhile, the attack is stealthy, achieving 98.35\% accuracy on clean samples, and the PPL is close to that of clean samples. 

\end{document}